\newif\ifarxiv
\arxivtrue

\RequirePackage{pdf14} 

\ifarxiv
    \documentclass[aps,prl,reprint, groupedaddress, nofootinbib]{revtex4-1}
    \usepackage{amsmath}
    \usepackage{standalone}
    \newcommand{\figurewidth}{0.45\textwidth}
    \renewcommand{\case}{\tfrac}
\else
    \documentclass[12pt]{iopart}
    \usepackage{iopams}
    \newcommand{\figurewidth}{0.6\textwidth}
    \newenvironment{align}{\begin{eqnarray}}{\end{eqnarray}}
\fi

\usepackage{graphicx}
\usepackage{braket}
\usepackage{upgreek}

\begin{document}

\ifarxiv
    \title{A Monte Carlo wavefunction method for semiclassical simulations of spin--position entanglement}
    \author{C.~J.~Billington}\email[]{chris.billington@monash.edu}
    \author{C.~J.~Watkins}
    \author{R.~P.~Anderson}
    \author{L.~D.~Turner}
    \affiliation{School of Physics and Astronomy, Monash University, Clayton, Victoria 3168, Australia}
    \date{\today}
\else

    \title[A MCWF method for semiclassical simulations of spin--position entanglement]
          {A Monte Carlo wavefunction method for semiclassical simulations of spin--position entanglement}

    \author{C~J~Billington, C~J~Watkins, R~P~Anderson and L~D~Turner}
    \address{School of Physics and Astronomy, Monash University, Clayton, Victoria 3168, Australia}
    \ead{chris.billington@monash.edu}
\fi

\begin{abstract}
We present a Monte Carlo wavefunction method for semiclassically modeling spin-$\frac12$ systems in a magnetic field gradient in one dimension. Our model resolves the conflict of determining what classical force an atom should be subjected to when it is in an arbitrary superposition of internal states. Spatial degrees of freedom are considered to be an environment, entanglement with which decoheres the internal states.
Atoms follow classical trajectories through space, punctuated by probabilistic jumps between spin states.

We modify the conventional Monte Carlo wavefunction method to jump between states when population transfer occurs, rather than when population is later discarded via exponential decay.
This results in a spinor wavefunction that is continuous in time, and allows us to model the classical particle trajectories (evolution of the environment variables) more accurately.
The model is not computationally demanding and it agrees well with simulations of the full spatial wavefunction of an atom.

\end{abstract}

\ifarxiv
\else

    \submitto{\NJP}
\fi

\maketitle

\section{Introduction}
A semiclassical simulation is one in which some degrees of freedom of a system are treated quantum mechanically and some classically.

Semiclassical simulations in which the positions and momenta of atoms are treated classically enjoy widespread use in cold atom physics, for example in laser cooling~\cite{stenholm1986, javanainen1992, dalibard89} and magnetic trapping/evaporative cooling~\cite{domokos2001}.
The benefit of not simulating the positions and momenta quantum mechanically is obvious: one need not spend the considerable computational resources required to do so. In addition, methods that define explicit classical trajectories are readily amenable to simulating collisions amongst ensembles of spins in magnetic fields using molecular dynamical methods such as direct simulation Monte Carlo (DSMC)~\cite{Wade2011}.

The downside is that superpositions of (and hence interference between) position and momentum states cannot be included in such a simulation.
This is usually not a problem, as the position evolution of an atom in a laser cooling simulation for example is well approximated by the action of a classical force---which doesn't result in superpositions of position states.

Furthermore, although atomic wavepackets tend to expand over time\footnote{For a thermal gas, they expand at a rate equal to the thermal velocity~\cite{busse2009}, which for a room temperature rubidium atom results in a wavepacket the size of a grapefruit after one millisecond!}, incoherent atomic scattering results in frequent position measurements such that atoms in a thermal cloud can be described by Gaussian wavepackets of a well defined average size that depends on the collective properties of the cloud~\cite{busse2009}.

So it would seem that such semiclassical simulations are ideal for cold atom simulations.
When are they not? Semiclassical simulations run into problems when it isn't apparent what classical force to use.
If an atom is in a superposition of two states with different magnetic moments in a magnetic field gradient, what force should it be subjected to? Corresponding to which magnetic moment?
The answer is of course both, and an actual atom in such a situation will evolve into a spin--position entangled state.

Since we wish to continue using semiclassical methods, what can we do about this? One approach is to use the \emph{average} force that the atom should feel, and this will indeed result in the correct expectation value of position over time (Ehrenfest's theorem~\cite{Ehrenfest1927}).
This works well when the atoms are mostly in one eigenstate, for example in Sisyphus cooling~\cite{dalibard89, salomon1990} where superpositions are dominated by groundstate population due to far detuned lasers and are short lived due to spontaneous emission (which puts atoms into eigenstates once more).

This approach does not work however when there are non-negligible and long-lived superpositions between states that see different potentials, which brings us to our motivation for developing the method that is the subject of this paper: forced evaporative cooling of neutral atoms in a magnetic trap.

In a magnetic field, the internal energy eigenstates of an atom are the states with well-defined spin projection in the local direction of the magnetic field, $\mathbf n  = \frac{\mathbf B}{|\mathbf B|}$. Atoms that are spin-aligned with the local field are attracted to regions of low field strength, and atoms anti-aligned are attracted to high field strength (or the reverse if the atom has a negative Land\'e $g$-factor). As such, atoms can be trapped in magnetic field configurations that have a spatial minimum of magnetic field strength somewhere, such as the quadrupole trap~\cite{pritchard1983} often used in Bose--Einstein condensation~\cite{davis1995, anderson1995}.

However, if the magnetic field direction is varying in space, these locally spin aligned or anti aligned states are \emph{not} eigenstates of the Hamiltonian an atom sees as it moves. If the atoms are moving slowly or the rate of change of the field direction is small, then the spins adiabatically follow the field and remain in a local eigenstate. However if they move more quickly, or though a region of rapidly changing field direction, transitions between local eigenstates can occur, called Majorana transitions~\cite{majorana1932}. This leads to atoms that were previously trapped becoming untrapped, and to losses in magnetic traps. Such Majorana losses are a problem in cold atom experiments, as cold atoms spend more time near the center of the trap where field directions change rapidly. This leads to both atom losses and overall heating of the atom cloud.

Using Ehrenfest's theorem leads to qualitatively incorrect results in this situation. An atom, having passed near to the field zero and transitioned into say, a 60:40 superposition of trapped and anti-trapped states, will feel a weakly trapping force according to Ehrenfest's theorem. It will not lose as much kinetic energy departing the field zero as it gained approaching it, and will eventually pass this kinetic energy on to other atoms in collisions.

An actual atom in this situation will of course diverge into two trajectories. The component of the superposition in the trapped spin state will remain on as tight an orbit about the field zero as before, not gaining any net energy from its close encounter. The anti-trapped component, conversely, will accelerate away from the field zero, and at typical evaporative cooling densities not collide with any other atoms on the way out.

The archetypal example of spin--position entanglement is the Stern--Gerlach experiment~\cite{Gerlach1922}, in which spin-$\frac12$ atoms---initially in an equal superposition of two spin projection states---traverse a magnetic field gradient.
Two distinct trajectories are observed resulting from the \emph{two distinct} forces on the two spin projection states. The expectation value of the force, however, is zero, and so simulating the Stern--Gerlach experiment with Ehrenfest's theorem similarly results in the qualitatively incorrect result of a single trajectory, in the center of where the two distinct trajectories would be.

\begin{figure}[t]
\begin{centering}
    \includegraphics[width=\figurewidth]{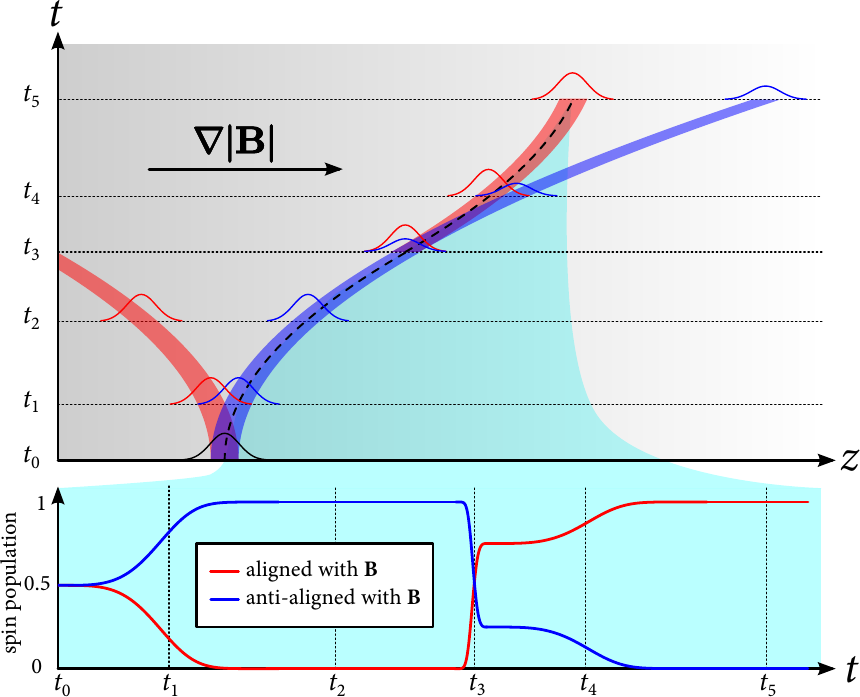}
    \caption{
    (Top) Single classical trajectory (dashed line) and spinor wavepacket extent (red and blue shaded).
    (Bottom) Spin populations (projected on local ${\bf B}(z)$) along classical trajectory.
     This schematic not to scale. Five relevant points in time are labeled. $t_0$: The atom is in a 50:50 superposition of locally spin up and spin down states. The spin up and spin down wavepackets accelerate away from each other. $t_1$: At the center of the spin down wavepacket, we now see a reduced spin up population, due to the two spin components separating. $t_2$: After sufficient separation time, we see no spin up population. $t_3$: The magnetic field vector changes direction rapidly, putting the atom into a 70:30 superposition of spin up and spin down. Our model chooses with 70\% probability  whether or not to start following the classical trajectory corresponding to the center of a spin up wavepacket. In this example it chooses to do so. $t_4$: The two wavepackets once again separate, and from the center of the spin up wavepacket, we see a reduced spin down population. $t_5$: After sufficient time we see only spin up population remaining. Note: In this schematic we've shown the decay in wavepacket overlap similarly to how it would actually look in reality: decreasing slowly at first, and then more rapidly as the wavepackets accelerate away from each other. However in our model we approximate simple exponential decay, and so the curves between $t_0$ and $t_1$, and between $t_3$ and $t_4$ are instead exponential in shape. \label{fig:schematic}}
 \end{centering}
\end{figure}

We solve this problem using a method in which there is only one classical force (corresponding to a single spin projection state) on each atom at each moment in time, despite the possibility of arbitrary superpositions of internal states.
The method can be visualized by considering two spin wavepackets separating as per the spatial Schr\"odinger equation---with the spin components accelerating in opposite directions---and asking ``What spin-state populations do I see if I follow the spin up wavepacket?".
As shown in figure~\ref{fig:schematic}, one sees the spin down population decreasing over time until only the spin up population remains. The rate that it does so depends on how fast the wavepackets separate. Likewise, if one follows the spin down component, one sees the spin up population decrease to zero. Our method does just this, following the trajectory of one spin component and decaying the other.

Although atoms mid-spin-flip during evaporative cooling do indeed evolve into spin--position entangled states, once separated, we are nonetheless happy to neglect possible future interference between these states. Firstly, the probability of the two wavepackets coming together again is low, and secondly, collisions with other atoms will soon occur, entangling atomic positions and decreasing the probability of future wavefunction overlap exponentially in the number of collisions.
Lastly, even if spatial interference does occur, it has little effect on our simulation. Interference fringes will be apparent on length scales close to the thermal wavelength, on which scale moving atoms around slightly won't affect the collective properties of the cloud. For these reasons, classical positions and momenta are suitable.

In the next section, we detail how our model is implemented, and compare it to more conventional MCWF methods.

\section{Spin decoherence from position separation}

Rather than computing the entire density matrix at every point in time, Monte-Carlo wavefunction (MCWF) methods~\cite{Carmichael1993, molmer1993, teich1992, gisin1992, wiseman1996, plenio1998} track a pure quantum state represented by a wavefunction.
This is computationally simpler, and results in the same statistical outcomes as the full density matrix approach.
It also has the appeal that pure quantum states are what we actually get when we perform quantum measurements, so every run of a MCWF simulation corresponds to an actual possible experimental outcome~\cite{Bouwmeester1996}.

Decoherence~\cite{zurek1991, zurek2003, schlosshauer2007} is introduced in the MCWF method by decaying some components of the wavefunction and not others~\cite{molmer1993, teich1992}.
Which basis this is done in depends on the system--environment interaction, and which states to decay and which to keep is probabilistic.
This arises from considering certain environmental states to be `classical', in the sense that one can ignore interference effects between them. This is usually valid due to the large number of degrees of freedom in environments. For example, the MCWF method is most often applied to decay of excited atomic states via spontaneous photon emission~\cite{Raithel2006}. The number of photons in the emitting mode is considered to be well defined at all times, as if a projective measurement of the photon number were being repeatedly performed~\cite{wiseman1996}.

Spin decoherence in our system arises from the fact that different spin states are subject to different spatial potentials, and hence spatially separate over time. The spin degree of freedom becomes entangled with the positional degree of freedom, and future interference between different spin states becomes unlikely. Our model approximates this process as irreversible, which allows us to model it similarly to environmentally induced decoherence.
This spatial separation results in a decreasing overlap between the wavepackets of the two spin components with time, decreasing the contrast of future interference between the spin populations, eventually to zero when they are completely separated.

This decrease in wavepacket overlap does not happen at a constant exponential rate, due to the Gaussian wavepacket shape and the fact that spin components accelerate away from each other.
In order to approximate the resulting decoherence as exponential decay, we compute an \emph{average} time $\tau$ taken for wavepackets to separate, and use its reciprocal as the decoherence rate.
This is the Markovian approximation often made in decoherence models~\cite{walls1985, moy1999}.
It comprises the assumption that the environment has no memory; in our case that our simulation will not keep a record of how long wavepackets have been accelerating apart.
The calculation of $\tau$ is detailed below.

In the language of decoherence, we have a \emph{system} to be modeled quantum mechanically---the spin state of an atom, and an \emph{environment} to be modeled classically---the position of said atom.
Our `classical' environmental states will be minimum-uncertainty Gaussian wavepackets~\cite{busse2009}. For a spin $F$ atom there are $2F+1$ such states---one for each spin projection eigenstate onto the local magnetic field.

Any interaction between a system and an environment will tend to diagonalize the system's reduced density matrix in the eigenbasis of the system-environment interaction Hamiltonian~\cite{zurek2003}, which for an atom in a magnetic field is the Zeeman Hamiltonian:
\begin{align}
\hat H_{\rm int} & = -\hat{\boldsymbol{\upmu}} \cdot \mathbf B(\hat{\mathbf r}),
\end{align}
with eigenvalues:
\begin{align}
E_{m_{\mathbf{n}}} = g_F m_{\mathbf{n}} \mu_{\rm B} |\mathbf B(\hat{\mathbf r})|,
\end{align}
where $m_{\mathbf{n}}$ is the spin projection quantum number in the direction of $\mathbf B$, $\hat{\mathbf r}$ is the position operator, $g_F$ is the atom's Land\'e $g$-factor for the states with total spin quantum number $F$, and $\mu_{\rm B}$ is the Bohr magneton. The eigenstates of this Hamiltonian are spin--position entangled states: it acts to separate spin states spatially so that they cannot interfere, and it will do so in the spin basis of the local magnetic field. As such, this is the basis in which we will be decohering states with respect to each other.

It might seem odd to consider the position degree of freedom of an atom to be an environment, capable of performing measurements on a system.
However for an interaction to appear as measurement-induced decoherence, it need only cause entanglement that makes future interference between the states that it measures negligible.
Quantum erasure experiments for example~\cite{Walborn2002} do not illustrate such interactions because the entanglement is reversible: interference can be restored. Entanglement that is near-complete and irreversible however, can be rightly regarded as strong measurement~\cite{zurek2003}.

For full, large scale environmental decoherence, such interference is unobservable due to the sheer number of degrees of freedom in large environments; the probability of two fragments of amplitude sharing a common origin ever encountering each other again in Hilbert space is vanishingly small~\cite{zurek2003}.

For us the probability is still small, but not nearly as vanishingly so.
However we are happy to make the approximation: that two of our Gaussian wavepackets, once separated, will not coincide in phase space at any time in the future prior to a collision (at which point full large scale decoherence with the rest of the cloud makes our approximation near exact).

Take a spin-$F$ system in a magnetic field gradient, for which spin and position are entangled. Its wavefunction is given by:
\begin{equation}
\ket\Psi = \sum_{m_{\mathbf{n}} = -F}^F
c_{m_{\mathbf{n}}} \ket{\psi_{m_{\mathbf{n}}}} \otimes \ket{m_{\mathbf{n}}},
\end{equation}
where $m_{\mathbf{n}}$ is the spin projection quantum number along the local magnetic field, and $\braket {\mathbf{r} |\psi_{m_{\mathbf{n}}}} = \psi_{m_{\mathbf{n}}}(\mathbf r)$ is the spatial wavefunction for that spin state (taken to be a Gaussian wavepacket following a classical trajectory).
The total density matrix for a spin-$\frac12$ system is therefore:
\begin{equation}
\rho(\mathbf r) = \left[\begin{array}{cc}
|c_\uparrow|^2|\psi_\uparrow|^2 & c_\uparrow c^*_\downarrow\psi_\uparrow\psi^*_\downarrow \\
\\
c_\downarrow c^*_\uparrow\psi_\downarrow\psi^*_\uparrow & |c_\downarrow|^2|\psi_\downarrow|^2\\
\end{array}\right],
\end{equation}
where the $\psi_{m_{\mathbf{n}}}(\mathbf r)$ are written as $\psi_\uparrow$ and $\psi_\downarrow$ for brevity.
This is the density matrix for the entire system plus environment we are considering, and so is a pure density matrix. If we now do a partial trace over the environmental degrees of freedom~\cite{schlosshauer2007}, we get the (possibly mixed) reduced density matrix for the atom's internal state:
\begin{equation}
\rho_{\rm spin} = \int \rho(\mathbf r)\,\mathrm{d}\mathbf{r} =
\left[\begin{array}{cc}
|c_\uparrow|^2 & c_\uparrow c^*_\downarrow\braket{\psi_\uparrow|\psi_\downarrow} \\
\\
c_\downarrow c^*_\uparrow\braket{\psi_\downarrow|\psi_\uparrow} & |c_\downarrow|^2  \\
\end{array}\right].
\end{equation}

Now consider an atom that starts at $t=0$ in a superposition of spin states, but with no spin--position entanglement---all the spatial wavefunctions $\psi_{m_{\mathbf{n}}}$ (which are really functions of time too, as we will soon see) are equal.
So each of the inner products above is equal to unity and we have a pure density matrix for our spin degree of freedom.

As time elapses, the spatial wavefunctions, each corresponding to a different spin state, begin to separate, moving in different directions. The center of each wavepacket will move with the classical force:
\begin{align}
{\mathbf F_{m_{\mathbf{n}}}} & = -\nabla(-{\boldsymbol{\upmu}}\cdot{\mathbf B})\\
                             & = -g_F m_{\mathbf{n}} \mu_{\rm B} \nabla|{\mathbf B}|.
\label{eq:force}
\end{align}
We model these spatial wavepackets as Gaussians with a fixed width.
The wavepacket overlap decreases with time as they spatially separate\footnote{See supplementary material for details.}:
\begin{align}
\left|\braket {\psi_i (t)| \psi_j (t)}\right| & =
\exp\left[
- \frac1{32\sigma^2}a_{ij}^2 t^4 - \frac{m^2\sigma^2}{2\hbar^2} a_{ij}^2 t^2
\right],
\label{eq:overlapdecrease}
\end{align}
where $m$ is the atom's mass, $a_{ij}$ is the magnitude of the relative acceleration between the two wavepackets as computed from the force (\ref{eq:force}), and $\sigma$ is the size of the Gaussian wavepackets. This is not the simple exponential decay required by the Markovian approximation. We seek an \emph{average} rate of separation $\tau_{ij}$ instead, such that the overlap decreases as:

\begin{equation}
\left|\braket{\psi_i (t)|\psi_j (t)}\right| = \exp\left[{-\frac t {\tau_{ij}}}\right].
\label{eq:markovian}
\end{equation}
Using (\ref{eq:overlapdecrease}) as the probability that the wavepackets are still overlapped, we may compute the expectation value of the time they are no longer overlapped. This is what we will use as the time constant $\tau_{ij}$ in (\ref{eq:markovian}):
\begin{align}
\tau_{ij} = \braket{t}_{ij} &= \frac{\int_0^\infty t \left|\braket{\psi_i (t)|\psi_j (t)}\right|\,\mathrm{d}t}
                                            {\int_0^\infty \left|\braket{\psi_i (t)|\psi_j (t)}\right|\,\mathrm{d}t}.
\end{align}
This gives us:
\begin{align}
\tau_{ij\,\mathrm{exact}} & =
\frac{
\sqrt{2\pi}\hbar\exp\left[\eta^2\right]\mathrm{erfc}\left(\sqrt{2}\eta\right)
}
{
m\sigma a_{ij}K_{\frac14}\left(\eta^2\right)
},
\label{eq:tauexact}
\end{align}
where $\mathrm{erfc}$ is the complementary error function, $K_{\frac14}$ is a modified Bessel function of the second kind and \mbox{$\eta = m^2\sigma^3a_{ij}\hbar^{-2}$} is a dimensionless parameter.

This is quite a mouthful, and is numerically difficult to evaluate. However, in the small wavepacket limit
$\eta \ll 1$, in which position separation dominates (\ref{eq:overlapdecrease}), one obtains:
\begin{align}
\tau_{{ij}\, \mathrm{pos}} & = \frac{(2\pi^2)^{\frac14}}{2\Gamma(\frac54)}
                               \sqrt{\frac\sigma{a_{ij}}}\\\nonumber\\
                            & \approx 1.163\sqrt{\frac\sigma{a_{ij}}}.
\label{eq:tau_pos}
\end{align}
In the large wavepacket limit $\eta \gg 1$, in which velocity separation dominates, one instead obtains:
\begin{align}
\tau_{{ij}\, \mathrm{vel}} & = \sqrt{\frac2\pi}\frac\hbar{m\sigma a_{ij}},
\label{eq:tau_vel}
\end{align}
and the following expression:
\begin{align}
\frac1{\tau_{ij\,\mathrm{approx}}} & = \left[\frac1{\tau^3_{ij\,\mathrm{pos}}} + \frac1{\tau^3_{ij\,\mathrm{vel}}}\right]^{\frac13},
\label{eq:tauapprox}
\end{align}
gives a good approximation to (\ref{eq:tauexact}) for all wavepacket sizes $\sigma$. We have studied the impact of using different values of $\tau$ and found that our method is relatively insensitive to the precise value calculated here.

We now have a set of \emph{decoherence times} $\tau_{ij}$ between each pair of states. Note that these are functions of the local magnetic field gradient and temperature---they should be recalculated appropriately for each atom as the simulation proceeds.

These times describe how long a spin superposition can live before the two spin states spatially separate and thus can no longer interfere.
As such, in a short time $\Delta t$, our pure reduced density matrix will become slightly mixed:
\begin{equation}
\rho_{\rm spin}(\Delta t) = \left[\begin{array}{cc}
|c_\uparrow|^2 & c_\uparrow c^*_\downarrow e^{-\frac{\Delta t}{\tau}} \\
\\
c_\downarrow c^*_\uparrow e^{-\frac{\Delta t}{\tau}} & |c_\downarrow|^2  \\
\end{array}\right],
\label{eq:decaydensitymatrix}
\end{equation}
where $\tau = \tau_{\uparrow\downarrow} = \tau_{\downarrow\uparrow}$ (for a spin-$\frac12$ system there is only one spin decoherence time).
In this way one obtains a reduced density matrix with decaying off diagonals, telling us that this is the basis in which spin decoherence occurs, and at what rate it does so.

Conditioned on a particular outcome of a spin measurement, we can then make statements about what the wavefunction looks like over time. If the particle is eventually determined by measurement to be spin up, then the wavefunction over time will have the spin down component decaying to zero. This is equivalent to decomposing the reduced density matrix at each moment in time into the sum of pure density matrices:

\begin{align}
\rho_{\rm spin}(\Delta t) &= \left[\begin{array}{cc}
|c_\uparrow|^2 & c_\uparrow c^*_\downarrow e^{-\frac{\Delta t}{\tau}} \\
\\
c_\downarrow c^*_\uparrow e^{-\frac{\Delta t}{\tau}} & e^{-2\frac{\Delta t}{\tau}}|c_\downarrow|^2   \\
\end{array}\right]
\ifarxiv
    \nonumber\\
\fi
&+\left[\begin{array}{cc}
0 & 0 \\
\\
0 & (1-e^{-2\frac{\Delta t}{\tau}})|c_\downarrow|^2  \\
\end{array}\right],
\label{eq:densitymatrixdecompose}
\end{align}
and discarding the second term each time. The conventional MCWF method allows for the possibility of instead discarding the first term and keeping the second one, resulting in an instantaneous change to the pure wavefunction being simulated; a `quantum jump'. At each timestep it chooses between the two above terms, weighted by the squared amplitudes of their corresponding wavefunctions. In our method, however, we have already decided in advance that we will take the first term, based on earlier population transfer between the two spin states. Alternately, a simulation with our method may begin, based on population transfer during some integration timestep, to start `tracking' the spin down state, after which the decomposition is instead:

\begin{align}
\rho_{\rm spin}(\Delta t) &= \left[\begin{array}{cc}
e^{-2\frac{\Delta t}{\tau}}|c_\uparrow|^2 & c_\uparrow c^*_\downarrow e^{-\frac{\Delta t}{\tau}} \\
\\
c_\downarrow c^*_\uparrow e^{-\frac{\Delta t}{\tau}} & |c_\downarrow|^2   \\
\end{array}\right]
\ifarxiv
    \nonumber\\
\fi
&+\left[\begin{array}{cc}
(1-e^{-2\frac{\Delta t}{\tau}})|c_\uparrow|^2 & 0 \\
\\
0 & 0  \\
\end{array}\right],
\end{align}
once again discarding the second term at each timestep. Our method of deciding in advance which term to discard, described in the next section, results in correct statistical outcomes for spin measurements (as shown in the results section by comparison with the spatial Schr\"odinger equation), and is in that regard identical to the conventional MCWF method. However, by bringing forward the decision as to which term of the density matrix decomposition to discard, we avoid instantaneous changes in the wavefunction, and more accurately simulate the classical trajectory by acknowledging a change in the spin state sooner.

\section{Method}
Here we present our model in one dimension and for a spin-half atom in one dimension.

In our model, each atom's state is fully described by its position $z$, its velocity $v_z$, its locally spin up and spin down populations $c_\uparrow$ and $c_\downarrow$, and its \emph{currently tracked state}, that is, which of the internal states $\ket\uparrow$ or $\ket\downarrow$ we are using to compute the classical force on the atom at that moment in time. We use language like `we are tracking the $\ket\uparrow$ component'. The currently tracked state can be stored simply as an integer, for example $+1$ for locally spin up, and $-1$ for locally spin down. A \emph{spin flip}, in the context of this method, is when we change, in the course of a simulation, which of the two states we are tracking.

Below, whenever we mention without qualification \emph{spin up} and \emph{spin down}, $\ket\uparrow$ and $\ket\downarrow$, or their amplitudes $c_\uparrow$ and $c_\downarrow$; we are referring to states of well defined spin projection in the direction of the local magnetic field the atom in question sees, not in the $z$ direction or any other lab basis.

The model is based on tracking one spin component at a time, and assuming its spatial wavefunction is a fixed-size Gaussian wavepacket with mean position following a classical trajectory according to the potential that spin component experiences. As in the conventional MCWF method, we exponentially decay the other, untracked, spin state according to the decoherence time between the pair of states.

Unlike the conventional MCWF method, we do not wait until the simulation discards population via exponential decay to consider `jumping' to other states. In developing the present model, we observed that doing so results in correct spin flip probabilities, but slightly incorrect classical trajectories.
We attribute this to the time delay between population transfer taking place and a subsequent `jump' occurring in the conventional MCWF method, a delay of on average one exponential time constant.
The atom experiences the wrong classical force in the meantime, which leads to discrepancies between the resulting trajectories and those expected from full spatial Schr\"odinger equation simulations.

Instead, we determine spin flip probabilities based on population transfer between spin states during each timestep. As a result, we do not need to `jump' to another state by instantaneously changing the wavefunction when we decide to start tracking a different spin state. We simply change which state the simulation is exponentially decaying. This represents a significant departure from the jumps in the conventional MCWF method, the physicality and interpretation of which is the topic of more general ongoing debate and discussion~\cite{wiseman1996, gisin1992}.

The instantaneous jumps of the conventional MCWF method may be interpreted in the context of our method as accounting for antecedent population transfer (on average, one decoherence time ago).
While such simulations produce correct final results for the system described by the wavefunction, they do not allow correct simulation of the corresponding classical environmental states---which we are interested in too. To clarify, we have \emph{not} done away entirely with jumps in the non-environment variables, this would clearly be unphysical since such jumps are very real in many experiments~\cite{Bergquist1986, Sauter1986}. Our system wavefunction still `jumps' from one spin state to another, but it does so continuously over the decoherence time, instead of instantaneously. For short decoherence times, this nonetheless may appear as a practically instantaneous change in the system's wavefunction.

From the vector spin operator $\hat{\mathbf{S}} = \{\hat{S}_x, \hat{S}_y, \hat{S}_z\}$, we can construct a spin projection operator  $\hat{S}_{\mathbf{n}} =  \mathbf{n} \cdot \hat{\mathbf{S}}$ in a direction described by the unit vector $\mathbf{n}\ =\ \{n_x, n_y, n_z\}$, which for us is the direction of the local magnetic field at each atom's position.
For a spin-$\frac12$ system, the projection operators onto the eigenstates of $\hat{S}_{\mathbf{n}}$ are:

\begin{align}
\hat P_\uparrow & = \ket{\uparrow}\bra{\uparrow} \nonumber\\
           & = \ket{m_{\mathbf{n}}=+\case12}\bra{m_{\mathbf{n}}=+\case12},\nonumber\\
P_\uparrow &= \left[\begin{array}{cc}
\frac12(1+n_z) & \frac12(n_x-in_y)\\
\\
\frac12(n_x+in_y) & \frac12(1-n_z)
\end{array}\right],
\label{eq:Pup}
\end{align}
and:
\begin{align}
\hat P_\downarrow & = \ket{\downarrow}\bra{\downarrow} \nonumber\\
             & = \ket{m_{\mathbf{n}}=-\case12}\bra{m_{\mathbf{n}}=-\case12},\nonumber\\
P_\downarrow &= \left[\begin{array}{cc}
\frac12(1-n_z) & -\frac12(n_x-in_y)\\
\\
-\frac12(n_x+in_y) & \frac12(1+n_z)
\end{array}\right],
\label{eq:Pdown}
\end{align}
where $P_\uparrow$ and $P_\downarrow$ are the matrix representations of $\hat P_\uparrow$ and $\hat P_\downarrow$ in the $z$ spin basis.
These projection operators are used at various steps in the model to project the spinor wavefunction onto the eigenstates of spin projection onto the local magnetic field.

\subsection{Algorithm}

To simulate classical initial conditions, simply set the particles' positions and velocities accordingly.
For initial conditions intended to approximate those of a spatially extended wavepacket, independently draw positions and velocities randomly from the position and velocity probability distributions of the spatial wavefunction in question\footnote{If you require position-velocity correlations for your initial conditions, use an appropriate joint probability distribution for $z$ and $v_z$, the correct derivation of which is beyond the scope of this paper.}.

To simulate any initial spin superposition, set $c_\uparrow$ and $c_\downarrow$ for each atom accordingly, ensuring normalization is satisfied: $|c_\uparrow|^2 + |c_\downarrow|^2 = 1$. Compute the eigenvectors of the local spin operator $\hat{S}_{\mathbf{n}}$; $\ket\uparrow$ and $\ket\downarrow$  in the $z$ basis for each atom (or whichever basis you will be simulating in), and construct that atom's initial spinor wavefunction:
\begin{equation}
\ket{\chi} = c_\uparrow\ket{\uparrow} + c_\downarrow\ket{\downarrow}.
\end{equation}

For each atom, choose which state initially to track. Do so by choosing randomly from the locally spin up and spin down states, weighted by their populations $|c_\uparrow|^2$ and $|c_\downarrow|^2$. For example, if all $N$ atoms start in a 50:50 superposition of spin up and down, this should result in $ (N \pm \sqrt N)/ 2$ atoms initially being tracked as spin up, and the rest as spin down.

The below steps describe integration for one atom in the model.
Atoms evolve independently, and so the method is trivially parallelizable.

\begin{enumerate}
\item At the start of each integration step, note the state populations:
\begin{equation}
n_\uparrow(t) = |c_\uparrow(t)|^2 = \bra{\chi(t)}\hat P_\uparrow\ket{\chi(t)},
\end{equation}
 and:
 \begin{equation}
 n_\downarrow(t) = |c_\downarrow(t)|^2 = \bra{\chi(t)}\hat P_\downarrow\ket{\chi(t)}.
 \end{equation}
 \label{item:notepops}
\item Do ordinary Hamiltonian evolution of the spin state for one timestep $\Delta t$:
\begin{equation}
\ket{\chi(t + \Delta t)} = \exp\left(-\frac i\hbar \hat H \Delta t\right)\ket{\chi(t)},
\label{eq:hamiltonian_evolution}
\end{equation}
 where $\hat H$ is the Zeeman Hamiltonian ${-\hat{\boldsymbol{\upmu}}}\cdot \mathbf B(z)$. Any integration method can be used, this unitary, first order method is shown as an example.\label{item:hamiltonian_evolution}

\item Evolve the position and velocity for one timestep $\Delta t$ according to the classical force on the state being tracked:
\begin{align}
{\mathbf F_{m_{\mathbf{n}\,\rm tracked}}} & = -\nabla(-\hat{\boldsymbol{\upmu}}\cdot{\mathbf B(z)}) \nonumber \\
           & = -g_F m_{\mathbf{n}\,\rm tracked} \mu_{\rm B} \nabla|{\mathbf B(z)}|.
\end{align}
Depending on the integration method, this might be done simultaneously with step~\ref{item:hamiltonian_evolution}.\label{item:classical_evolution}

\item Note the state populations again, and determine from the state populations noted in step~\ref{item:notepops}, how much probability moved from the state currently being tracked, to the other state during this integration timestep. Compute this as a fraction of the currently tracked state's population. For example, if the simulation is currently tracking the atom's spin up state, then we have:
\begin{equation}
p_{\rm flip} =
\frac{|c_\downarrow(t + \Delta t)|^2 - |c_\downarrow(t)|^2}{|c_\uparrow(t + \Delta t)|^2}.
\label{eq:poptransfer}
\end{equation}
\label{item:compute_transitions}
\item Draw a random number between zero and one. If it's less than $p_{\rm flip}$, then the atom is to undergo a spin flip, if doing so is energetically allowed. The above quantity $p_{\rm flip}$ might be less than zero, indicating probability flow the other way (from the untracked state to the tracked one), which is OK and will result in zero chance of a spin flip when the positive random number is chosen. If no spin flip is to occur, skip to step~\ref{item:expdecay}.

\item If a spin flip is to occur, compute the difference in potential energy between the currently tracked state and the other state, that is, by how much would an atom's potential energy change if it were to flip from the tracked state to the untracked one at this point in space?
\begin{align}
\Delta E_{\rm pot} = V_{\rm untracked}(z) - V_{\rm tracked}(z) \\
                        \qquad = g_F (m_{\mathbf{n}\,{\rm untracked}} - m_{\mathbf{n}\,{\rm tracked}}) \mu_{\rm B} |{\mathbf B(z)}|.
\end{align}
Compare with the atom's kinetic energy $E_{\rm kin}$. If $E_{\rm kin} + \Delta E_{\rm pot} < 0$, the spin flip is not (classically) energetically allowed. Continue tracking the original state and skip to step~\ref{item:expdecay}.

Otherwise, modify the magnitude of the atom's velocity $v_z$, but not the direction, so as to conserve energy:
\begin{equation}
E_{\rm kin} \rightarrow E_{\rm kin} + \Delta E_{\rm pot}
\end{equation}
\begin{equation}
\Rightarrow |v_z| \rightarrow \sqrt{\frac{2(E_{\rm kin} + \Delta E_{\rm pot})}m}.
\end{equation}

Finally, actually perform the spin flip---flipping entails simply noting that the other state is now being tracked, and does not involve modifying the wavefunction in any way. \label{item:velocitykicks}

\item Exponentially decay for one timestep the state not being tracked:
\begin{equation}
\ket{\chi} \rightarrow \left(1-\frac {\Delta t}\tau \hat P_{\rm untracked}\right)\ket{\chi},
\end{equation}
where $\hat P_{\rm untracked}$ , either $\hat P_\uparrow$ or $\hat P_\downarrow$, defined in (\ref{eq:Pup}) and (\ref{eq:Pdown}) is the projection operator onto the state not being tracked, and $\tau$, given in (\ref{eq:tauapprox}) is the spin decoherence time.
Again, the numerical method used for this integration step is not important, a simple Euler method step is shown here as an example, and was accurate enough for our examples, for which $\Delta t << \tau$.\label{item:expdecay}

\item Normalize the wavefunction:
\begin{equation}
\ket{\chi}\rightarrow \frac{\ket{\chi}}{\sqrt{\braket{\chi|\chi}}},
\end{equation}
and proceed to step~\ref{item:notepops}.
\end{enumerate}

\section{Results}\label{sec:results}

In this section we compare our method to two others: the full spatial time-dependent Schr\"odinger equation\footnote{Simulated using the XMDS~\cite{Dennis2013} differential equation solver.} (TDSE), and the standard semiclassical method (which we refer to here as the Ehrenfest method) mentioned in the introduction, whereby we calculate an expectation value for the force on the atom using the instantaneous spin populations of the atom. In this section `Monte Carlo wavefunction method' or MCWF refers to our method.

We first test our model on a simplified form of the Stern--Gerlach experiment, and subsequently test on a modified form of the Majorana problem which describes a 1D magnetic trap with the possibility of Majorana losses near the center.

The results shown for both the MCWF and Ehrenfest simulations are ensemble averages for simulations done with 10,000 atoms.
The simulations begin with a stationary, 20\,$\upmu$K $^{87}$Rb atom that has a minimum uncertainty wavepacket: the spatial and momentum distributions are Gaussian with widths $\lambda_{\mathrm{th}}$ and $\hbar/2\lambda_{\mathrm{th}}$ respectively, where $\lambda_{\mathrm{th}} = h/\sqrt{2\pi m k_{\mathrm{B}} T}$ is the thermal de Broglie wavelength.

Throughout this section we treat $^{87}$Rb as a spin-$\frac12$ atom with a Land\'e $g$-factor $g_F$ equal to twice that of $^{87}$Rb's actual spin-1 hyperfine groundstate.

Population densities (and populations) for the TDSE are computed simply as the populations (and integrals) of its two internal spin states, in the basis of the local magnetic field.
The same quantities for the MCWF method, however, are computed not by reference to the internal state of the atom, but instead by counting what proportion of the atoms are being tracked as spin up or spin down.

\subsection{The Stern--Gerlach experiment}

\begin{figure}[t]
    \begin{centering}
    \includegraphics[width=\figurewidth]{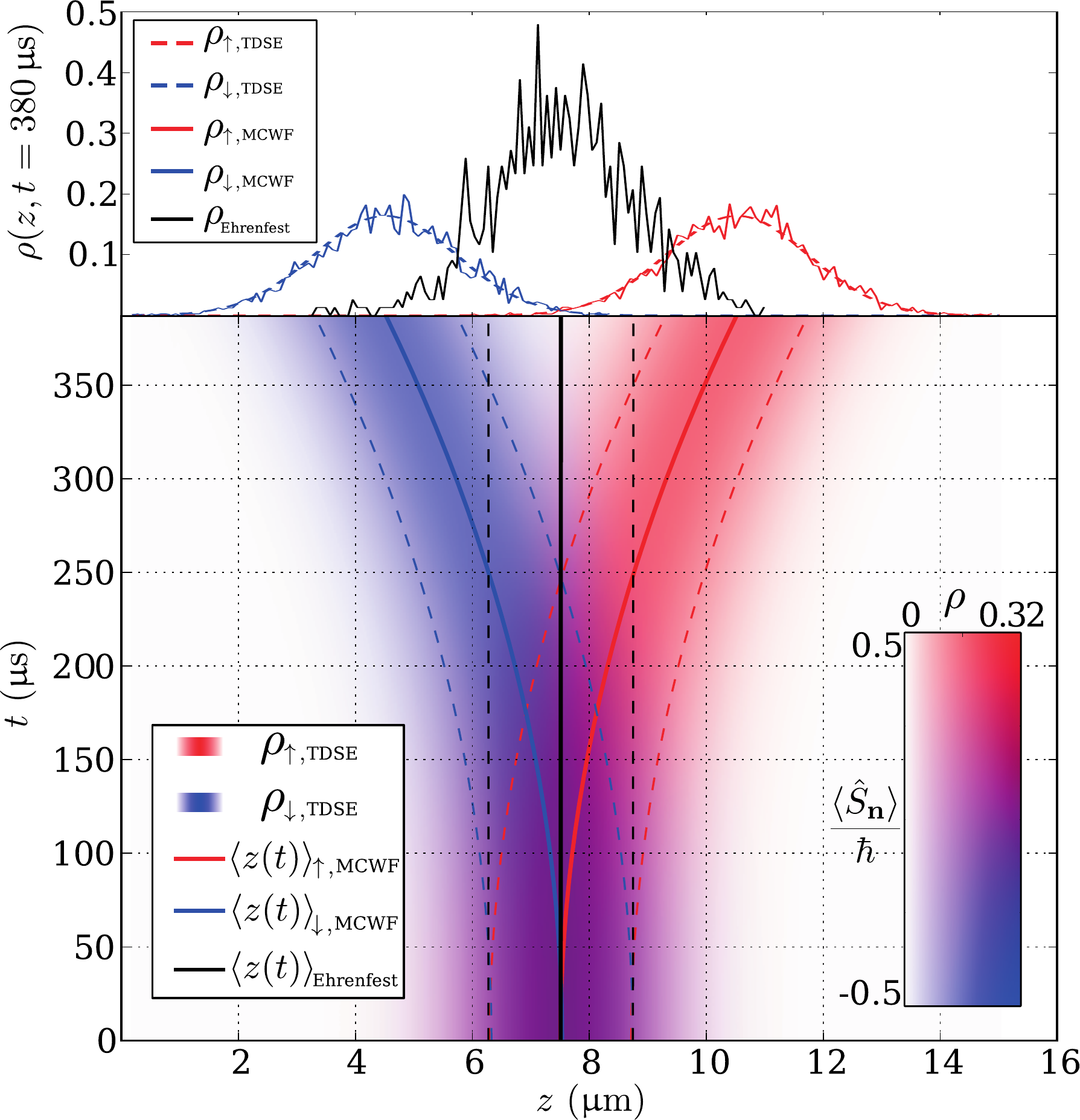}
    \caption{ Results of the MCWF, TDSE and Ehrenfest methods applied to a simple 1-dimensional Stern--Gerlach simulation.
    The shading shows the results of the TDSE simulation, the colored lines show the trajectories predicted by the MCWF method, and the black lines illustrate the trajectories predicted by the Ehrenfest method.
    Solid lines represent the mean trajectories, and dashed lines show one standard deviation from the mean.
    The plot at the top of the figure illustrates the probability densities predicted by all three methods at the end of the simulation ($t=380\,\upmu\mathrm{s}$).
    Here solid colored lines are the results of the MCWF simulation, dashed colored lines represent the predictions of the TDSE simulation, and the black line is the prediction of the Ehrenfest method.\label{fig:SternGerlach}}
\end{centering}
\end{figure}

As discussed in the introduction, a prototypical example of spin--position entanglement is the Stern--Gerlach experiment.
The original Stern--Gerlach experiment saw spin-half atoms passed through an inhomogeneous magnetic field and after some propagation two distinct atom clouds were seen, implying the passage of two distinct trajectories, corresponding to the two distinct spin projections along the direction of the field.
Our method captures the essential physics of the Stern--Gerlach experiment in one dimension by simulating a spin-half atom beginning at rest in an equal superposition within a magnetic field gradient along $z$.
Figure~\ref{fig:SternGerlach} shows the results predicted by our MCWF method contrasted with TDSE and Ehrenfest solutions.

Our initial wavepacket is centered at \mbox{$z=7.8\,\upmu$m} in the magnetic field \mbox{$\mathbf{B}(z)=(0,0,B_z^\prime z)$}, with \mbox{$B_z^\prime = 2.5\,\mathrm{T\,\mathrm{m}}^{-1}$}.
The wavepacket begins at rest, in a superposition of spin projection states ($c_{\uparrow}=c_{\downarrow}=1/\sqrt{2}$), which accelerate apart in the magnetic field gradient.
This example shows perhaps the simplest case in which using an expectation value for the force on an atom is completely inapplicable.
We see that the Ehrenfest approach results in the atoms remaining stationary, as one would expect given that the average of the forces the two components experience is zero.
The shaded density plot shows qualitative agreement between the MCWF and TDSE simulations, with the respective spin components tracking similar trajectories.
The inset shows the quantitative accuracy of the MCWF method, displaying good agreement between the density profile it predicts and that of the TDSE method at the final time step of the simulation.
The noise evident in both the Ehrenfest and MCWF simulations is statistical, and can be reduced by the addition of more simulation atoms.

\subsection{The Majorana problem}

Earlier we introduced the concept of a Majorana transition and discussed the detrimental effect it can have during an ultracold atom experiment.
In other work of ours, we want to model this effect using a full three dimensional gas simulation, modeling classical positions and momenta in order to scale to large numbers of atoms.

Majorana derived a result~\cite{majorana1932} predicting the probability of a transition based on the ratio between the rate of change of the magnetic field direction and the Larmor frequency.
However this derivation was made for a stationary atom in a dynamic field, which---unlike a moving atom in a \emph{spatially} varying field---does not develop spin--position entanglement.

Moving into the center of mass frame of a moving atom in a spatially varying field is not sufficient to make the situation equivalent to Majorana's assumptions, since he assumed a time-varying magnetic field only, with no spatial gradients that could give rise to forces.
It was also derived in the asymptotic limit of a large longitudinal field---initially aligned with spin---which is steadily inverted in the presence of a constant transverse field.
Nonetheless we consider a similar situation with a moving atom subject to forces in a magnetic field gradient, and compare the spin flip probabilities it predicts to those predicted by Majorana for the effectively time-dependent magnetic field the moving atom sees at the field minimum.

\begin{figure}[t]
\begin{centering}
    \includegraphics[width=\figurewidth]{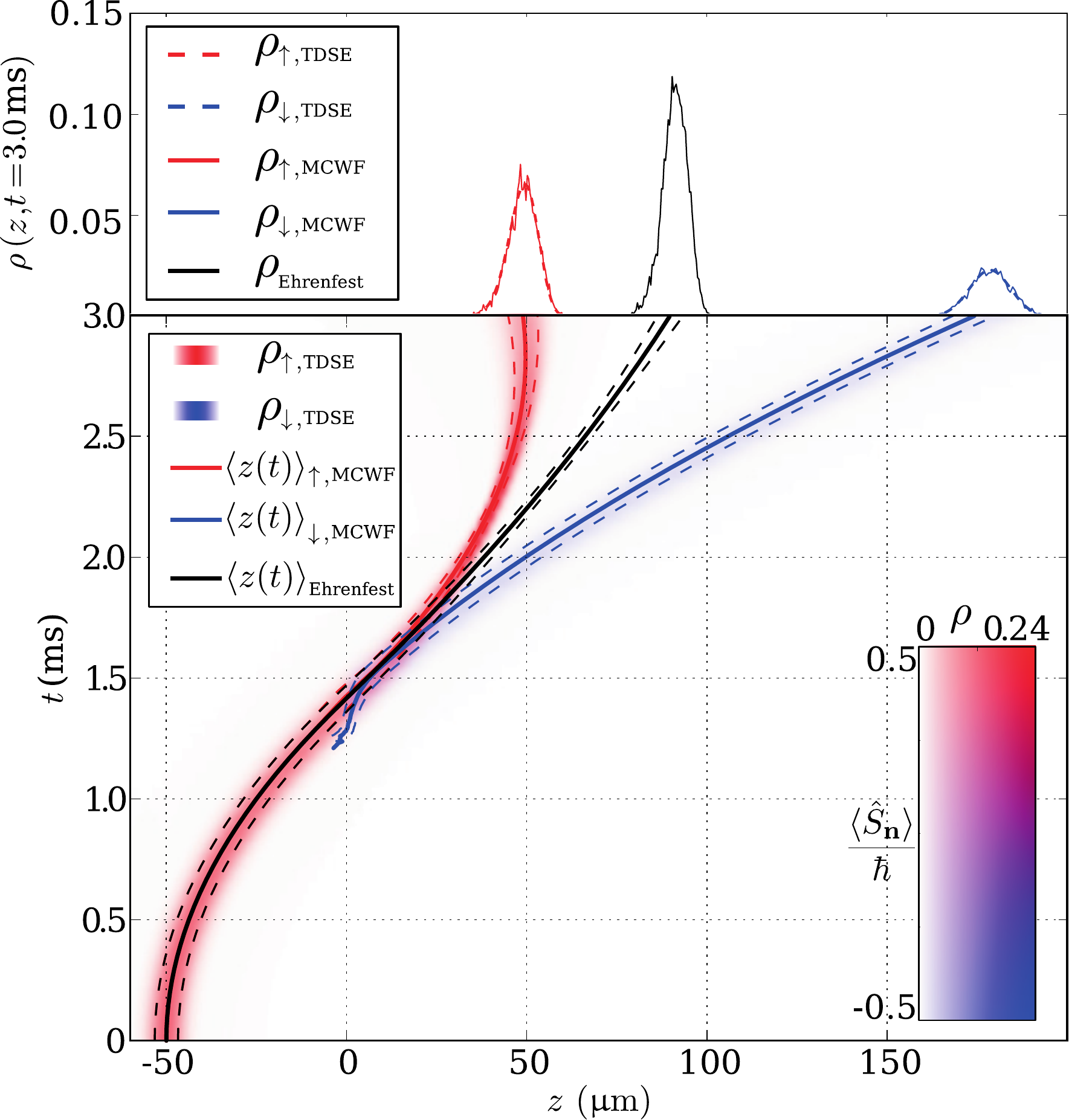}
    \caption{ The modified Majorana problem of an atom accelerating in a 1-dimensional magnetic field gradient, simulated using the three methods: MCWF, TDSE and Ehrenfest.
    The density plot at the top of the figure is for $t=3.0\,\mathrm{ms}$ (as shown by the dashed line).
    The legend is as described in figure~\ref{fig:SternGerlach}.
    \label{fig:mmContourInset}}
\end{centering}
\end{figure}

\begin{figure}[t]
\begin{centering}
    \includegraphics[width=\figurewidth]{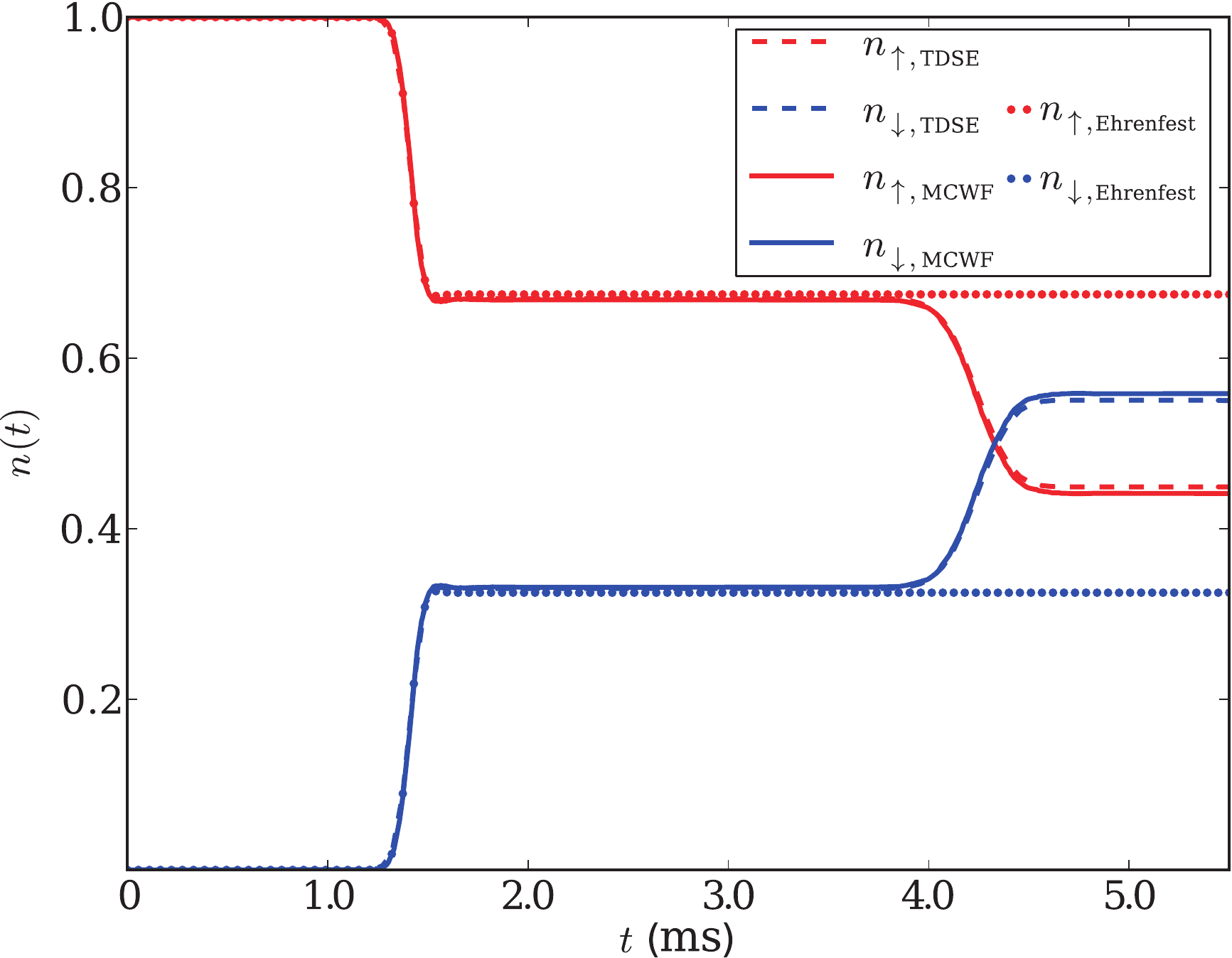} 
    \caption{ State population of the modified Majorana problem in figure~\ref{fig:mmContourInset}, here extended to $t=5.5\,\mathrm{ms}$.
    The Ehrenfest approximation correctly predicts populations for the first minimum crossing but subsequently fails for successive crossings.
    \label{fig:mmPops}}
\end{centering}
\end{figure}

We apply the MCWF method to a simulation of an atom moving in a magnetic field gradient along a single spatial dimension, the results of which are shown in Figures~\ref{fig:mmContourInset} and~\ref{fig:mmPops}.
The wavepacket is initially centered around \mbox{$z=-50\,\upmu$m} in the magnetic field \mbox{$\mathbf{B}(z)=(B_x,0,B_z^\prime z)$}, with \mbox{$B_x=105\,$nT} and \mbox{$B_z^\prime = 2.5\,$Tm$^{-1}$}.
In figure~\ref{fig:mmContourInset} we have again plotted the trajectories predicted by each method and provided an inset displaying the probability densities at the final time.

Figure~\ref{fig:mmPops} shows the evolution of the population in each spin state over time.
Here we show more simulation time to emphasize the failure of the Ehrenfest method.
Again we have compared the results of the three different methods; the TDSE, the MCWF and the Ehrenfest methods.
As before we note good agreement between the TDSE and the MCWF methods in the density plots, and observe the accuracy of the MCWF method in the density profiles plotted in the inset.
While the Ehrenfest trajectory initially agrees closely with the other two simulations, it diverges after the wavepacket has passed through the field minimum.
The Ehrenfest populations agree well through the first crossing of the minimum and it isn't until the trapped atom encounters another magnetic field minimum that we really observe the failure of the Ehrenfest method in simulating state populations.
This failure may be entirely attributed to the Ehrenfest method's manifest failure to correctly simulate the wavepacket trajectories---most of its atoms are on an escape trajectory and will not see a second crossing of the magnetic field minimum.
The stochastic nature of the MCWF method is evident in trajectory of the flipped state near the magnetic field minimum (figure~\ref{fig:mmContourInset}): here the population in the flipped state is small and so has large statistical noise. This noise decreases as the the flipped state population increases towards its asymptotic value.

\begin{figure}[t]
\begin{centering}
    \includegraphics[width=\figurewidth]{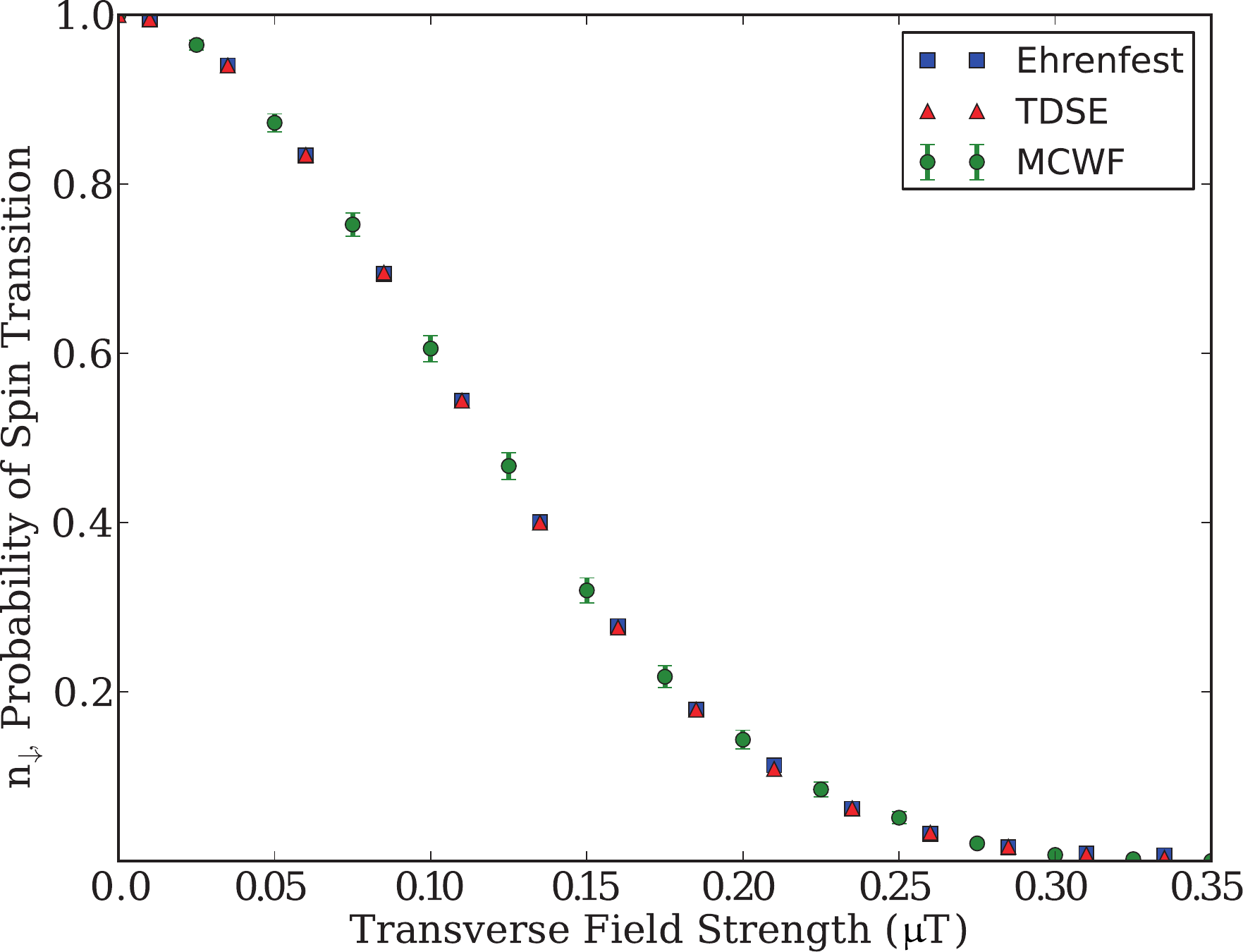} 
    \caption{ The modified Majorana simulations over a range of transverse magnetic fields tests the technique over a range of spin transition probabilities.\label{fig:varyingBx}}
\end{centering}
\end{figure}

To ensure robustness of the MCWF method over a practical range of spin flip probabilities, in figure~\ref{fig:varyingBx} we have plotted the predictions of the three methods over a range of transverse magnetic fields.
As the transverse field strength increases the rate of change of the magnetic field decreases and we expect to see a decrease in the probability of spin flip transitions.
Each point in figure~\ref{fig:varyingBx} is the final result of a simulation of the type shown in figure~\ref{fig:mmContourInset}.
These simulations all begin with a static 20\,$\upmu$K wavepacket of 10,000 $^{87}$Rb spin up atoms at $z=-29.8\,\upmu$m. This is far enough away from the field minimum to be considered asymptotic.
The atoms are allowed to accelerate through the magnetic field until they reach the antipodal point of the trapping potential.
As expected we see the MCWF method agrees within its statistical uncertainty with the TDSE method.
As before the Ehrenfest method produces good results, but if we were to allow the simulation to run longer we would soon see the method disagreeing with the MCWF and TDSE methods due to differing trajectories.

\section{Discussion}

\subsection{Computational costs}

The full many-body wavefunction is entirely infeasible to simulate for any more than a small number of atoms. Making the assumption that the atoms are in a product state allows on to simulate a single spatial Schr\"odinger equation per atom.
Compared to this, our method has the computational cost of one ODE per atom instead of one PDE, making it tractable for large numbers of atoms.

The results in
\ifarxiv
    the previous section 
\else
    section~\ref{sec:results}
\fi
do not demonstrate any such computational savings, since thousands of atoms were compared to a single TDSE run to demonstrate correctness. If one is content with only a single statistically representative outcome however, as is likely satisfactory for large thermal clouds of atoms, one need only perform one MCWF run. As in all stochastic methods, repeated runs can be used to establish the variability of results at the cost of more computing resources.

\subsection{Applicability}

Our method of determining spin flip probabilities from population transfer allows us to more accurately model the classical trajectories of the simulated atoms than if we waited for population to be discarded, as in existing MCWF methods. We stop short of claiming, however, that computing spin flip probabilities in this way is at all correct for simulating other types of open systems with quantum jump methods.

Indeed, it seems that if one were to try to apply this method to the common case of spontaneous emission of a photon from an atom, it would be the photon number states, not the atomic states, between which population transfer would have to be monitored. There is an extra layer present in this case compared to ours: the external environment is measuring photon number states, which in turn measure internal states of the atom. The time photon number states take to decohere, conditioned on photon detection with a large environment, is so short that even if one decided to use our method here, the post-jump exponential decay would be practically instantaneous and indistinguishable from the immediate jumps these methods use.

Another difference between our system and a decaying atom is that the positional states corresponding to each spin state are non-orthogonal (being mostly overlapping Gaussians), unlike photon number states. This means that the assumption of constant, strong position measurements does not produce complete collapse of the spinor wavefunction, as any given position measurement provides only weak spin information. It is therefore natural that our model does not produce discontinuous changes in the system wavefunction.

\subsection{Higher spins and higher spatial dimensions}

Generalizing our method to higher spins entails defining classical probability flows between each pair of states in a way that is consistent with the actual changes in population at each timestep. Inferring classical probability flows is trivial for a two state system, but more difficult for higher spins due to the inevitability of interference.

A recipe for inferring classical probability flows in a quantum system comprises a \emph{hidden-variables theory}, defined by Aaronson as ``a way to convert a unitary matrix that maps one quantum state to another into a stochastic matrix that maps the initial probability distribution to the final one in some fixed basis"~\cite{Aaronson2005}. According to this definition, our method is a hidden-variables theory, with the spin projection quantum number $m_{\mathbf{n}}$ being a hidden variable. Approaches to hidden-variables theories for discrete systems of more than two states exist, including the `flow' and `Schr\"odinger' theories~\cite{Aaronson2005}.

The only difficulty in generalizing to greater than one spatial dimension is knowing in which direction to modify atomic velocities when kinetic energy is lost or gained during a spin flip. One solution might be to calculate the instantaneous force on a moving classical dipole, and to apply a velocity difference to the atom in the direction of this force.
Another approach is to simply continue to apply velocity changes in the direction of motion in three dimensions. If the system is chaotic and there is no systematic bias making resulting trajectories statistically distinguishable from the correct ones, then such unphysical velocity adjustments may nonetheless be satisfactory.

\section{Conclusion}

We have presented a one dimensional model for semiclassical simulations of spin-$\frac12$ atoms in magnetic fields. Our model reproduces the salient features of the underlying exact model, namely correct spin flip probabilities and statistically representative trajectories, which the Ehrenfest method fails to do.

Our method is a great improvement over using Ehrenfest's theorem for force calculation in semiclassical simulations in contexts where superpositions of states subject to different forces are non-negligible and long-lived. We intend to apply the method to simulations of forced evaporative cooling, in the hopes of further closing the gap between theory and experiment on the subject.

\ifarxiv
    \begin{acknowledgments}
\else
    \ack
\fi

The particle simulations shown here were performed using a GeForce Titan donated by the NVIDIA Corporation.
The authors would like to thank Paul Lett and Kristian Helmerson for useful comments on our manuscript.

\ifarxiv
    \end{acknowledgments}
\else
    \section*{References}
    \bibliographystyle{unsrt}
\fi

\bibliography{bibliography}

\ifarxiv
    \onecolumngrid
    \setcounter{section}{0}
    \setcounter{equation}{0}
    \setcounter{footnote}{0}
    \setcounter{figure}{0}
    \clearpage
    \section*{Supplementary information:}
    \newif\ifarxiv
\arxivtrue
\documentclass[a4paper,11pt]{article}
\usepackage{amsmath}
\usepackage{braket}
\usepackage[top=3.5cm, bottom=4cm, left=3cm, right=3cm]{geometry}
\ifarxiv
    \usepackage{graphicx}
\else
    \usepackage{mathspec}
    \setmathfont(Digits,Latin)[Uppercase=Italic,Lowercase=Italic]{Minion Pro}
    \setmathfont(Greek)[Uppercase=Regular,Lowercase=Regular]{Minion Pro}
    \setmainfont[Mapping=tex-text, Contextuals=Alternate,Numbers=OldStyle]{Garamond Premier Pro}
    \expandafter\let\expandafter\hbar\csname ?-\string\hbar\endcsname
\fi

\begin{document}
\title{Supplementary Information to ``A Monte Carlo wavefunction method for semiclassical simulations of spin-position entanglement''}
\author{C~J~Billington\footnote{\texttt{chris.billington@monash.edu}}, C~J~Watkins, R~P~Anderson and L~D~Turner\\
School of Physics and Astronomy, Monash University, Clayton, Victoria 3168, Australia}

\maketitle
\section{Derivation of spin decoherence time $\tau$}
\subsection{Inner product of two Gaussian wavepackets accelerating apart}

We want to know what the inner product of two equal width Gaussian wavepackets $\psi_i$ and $\psi_j$ is as a function of time as they move apart with constant acceleration with magnitude $a_{ij}$:
\begin{align}
\braket {\psi_i(t) | \psi_j(t)} = C \int_{-\infty}^{\infty} e^{-\frac{x^2}{4\sigma^2}}e^{-\frac{(x-x_{\mathrm{rel}})^2}{4\sigma^2} + ik_{\mathrm{rel}}x}\,\mathrm{d}x,
\end{align}
where
\begin{align}
x_{\mathrm{rel}}(t) = \frac12a_{ij}t^2
\end{align}
and
\begin{align}
k_{\mathrm{rel}}(t) = \frac m \hbar a_{ij} t
\end{align}
are the wavepackets' relative position and wavenumber due to acceleration for a time $t$ starting from rest, and:
\begin{align}
C^{-1}=\int_{-\infty}^\infty e^{-\frac{x^2}{2\sigma^2}}\,\mathrm{d}x\label{supp:eq:Cdef}
\end{align}
is a normalization constant. Note that this expression holds for any number of dimensions --- relative motion is only along one axis so the integrals in all other directions equal one.

Let's evaluate this integral by expanding the whole exponent into a polynomial in x, and completing the square in the exponent:
\begin{align}
\braket {\psi_i(t) | \psi_j(t)} & = C \int_{-\infty}^{\infty} \exp\left[-\frac{x^2}{4\sigma^2} -\frac{(x-x_{\mathrm{rel}})^2}{4\sigma^2} + ik_{\mathrm{rel}}x\right]\,\mathrm{d}x
\\
& =  C\int_{-\infty}^{\infty} \exp\left[
-\frac{x^2}{4\sigma^2} -\frac{x^2}{4\sigma^2} + \frac{xx_{\mathrm{rel}}}{2\sigma^2} - \frac{x_{\mathrm{rel}}^2}{4\sigma^2} + ik_{\mathrm{rel}}x
\right]\,\mathrm{d}x
\\
& =  C\int_{-\infty}^{\infty} \exp\left[-\frac1{2\sigma^2}\left(
x^2 - xx_{\mathrm{rel}} - 2i\sigma^2 k_{\mathrm{rel}}x + \frac12 x_{\mathrm{rel}}^2
\right)\right]\,\mathrm{d}x
\\
& =  C\int_{-\infty}^{\infty} \exp\left[-\frac1{2\sigma^2}\left(
x^2 - \left[x_{\mathrm{rel}} - 2i\sigma^2 k_{\mathrm{rel}}\right]x + \frac12 x_{\mathrm{rel}}^2
\right)\right]\,\mathrm{d}x.
\end{align}
Now we've got a polynomial in $x$ in the exponent and can complete the square:
\begin{align}
p(x) & = x^2 - \left[x_{\mathrm{rel}} - 2i\sigma^2 k_{\mathrm{rel}}\right]x + \frac12 x_{\mathrm{rel}}^2
\\
& = x^2 - \left[x_{\mathrm{rel}} - 2i\sigma^2 k_{\mathrm{rel}}\right]x 
+ \frac14\left[x_{\mathrm{rel}} - 2i\sigma^2 k_{\mathrm{rel}}\right]^2
- \frac14\left[x_{\mathrm{rel}} - 2i\sigma^2 k_{\mathrm{rel}}\right]^2
+ \frac12 x_{\mathrm{rel}}^2
\\
& = \left(x - \frac12\left[x_{\mathrm{rel}} - 2i\sigma^2 k_{\mathrm{rel}}\right]\right)^2 
- \frac14\left[x_{\mathrm{rel}} - 2i\sigma^2 k_{\mathrm{rel}}\right]^2
+ \frac12 x_{\mathrm{rel}}^2
\end{align}

\begin{align}
\Rightarrow \braket {\psi_i (t)| \psi_j(t)} & = C\int_{-\infty}^{\infty} \exp\left[
-\frac{\left(x - \frac12\left[x_{\mathrm{rel}} - 2i\sigma^2 k_{\mathrm{rel}}\right]\right)^2}{2\sigma^2}
+ \frac1{8\sigma^2}\left[x_{\mathrm{rel}} - 2i\sigma^2 k_{\mathrm{rel}}\right]^2
- \frac1{4\sigma^2} x_{\mathrm{rel}}^2
\right]\,\mathrm{d}x
\\
& = \exp\left[
\frac1{8\sigma^2}\left[x_{\mathrm{rel}} - 2i\sigma^2 k_{\mathrm{rel}}\right]^2
- \frac1{4\sigma^2} x_{\mathrm{rel}}^2
\right]
C\int_{-\infty}^{\infty} e^{
-\frac{\left(x - c\right)^2}{2\sigma^2}}\,\mathrm{d}x
\end{align}
where $c$ is a complex number. We recognize the remaining integral as a Gaussian integral with complex offset --- it equals $C^{-1}$ as in (\ref{supp:eq:Cdef}). So it cancels $C$ and we're left with:
\begin{align}
\braket {\psi_i(t) | \psi_j(t)} & =
\exp\left[
\frac1{8\sigma^2}\left[x_{\mathrm{rel}} - 2i\sigma^2 k_{\mathrm{rel}}\right]^2
- \frac1{4\sigma^2} x_{\mathrm{rel}}^2
\right].
\end{align}
Expanding:
\begin{align}
\braket {\psi_i (t)| \psi_j(t)} & =
\exp\left[
\frac1{8\sigma^2}\left[x_{\mathrm{rel}}^2 - 4i\sigma^2 x_{\mathrm{rel}}k_{\mathrm{rel}} - 4\sigma^4 k_{\mathrm{rel}}^2\right]
- \frac1{4\sigma^2} x_{\mathrm{rel}}^2,
\right]
\end{align}
gives us our result in terms of $x_{\mathrm{rel}}$ and $k_{\mathrm{rel}}$:
\begin{align}
\braket {\psi_i(t) | \psi_j(t)} & =
\exp\left[
- \frac1{8\sigma^2}x_{\mathrm{rel}}^2 - \frac i2 x_{\mathrm{rel}}k_{\mathrm{rel}} - \frac{\sigma^2}2 k_{\mathrm{rel}}^2
\right]
\\
\Rightarrow \left|\braket {\psi_i(t) | \psi_j(t)}\right| & =
\exp\left[
- \frac1{8\sigma^2}x_{\mathrm{rel}}^2 - \frac{\sigma^2}2 k_{\mathrm{rel}}^2
\right].\label{supp:eq:modoverlap}
\end{align}
We see that for small wavepacket sizes $\sigma$, position separation dominates the decrease in wavepacket overlap, whereas for large wavepackets, velocity separation dominates. If it is known which of two terms in the exponent above dominates for a particular problem, here would be a good time to make the approximation of neglecting the smaller term. This will result in a simpler expression for the separation time $\tau_{ij}$ below.

Let's substitute back in the expressions for $x_{\mathrm{rel}}$ and $k_{\mathrm{rel}}$ to get a polynomial in $t$ in the exponent:
\begin{align}
\left|\braket {\psi_i(t) | \psi_j(t)}\right| & =
\exp\left[
- \frac1{32\sigma^2}a_{ij}^2 t^4 - \frac{m^2\sigma^2}{2\hbar^2} a_{ij}^2 t^2
\right].\label{supp:eq:timeconstants}
\end{align}
\subsection{Time of separation}

\begin{figure}[t]
    \begin{centering}
    \ifarxiv
        \includegraphics[width=0.7\textwidth]{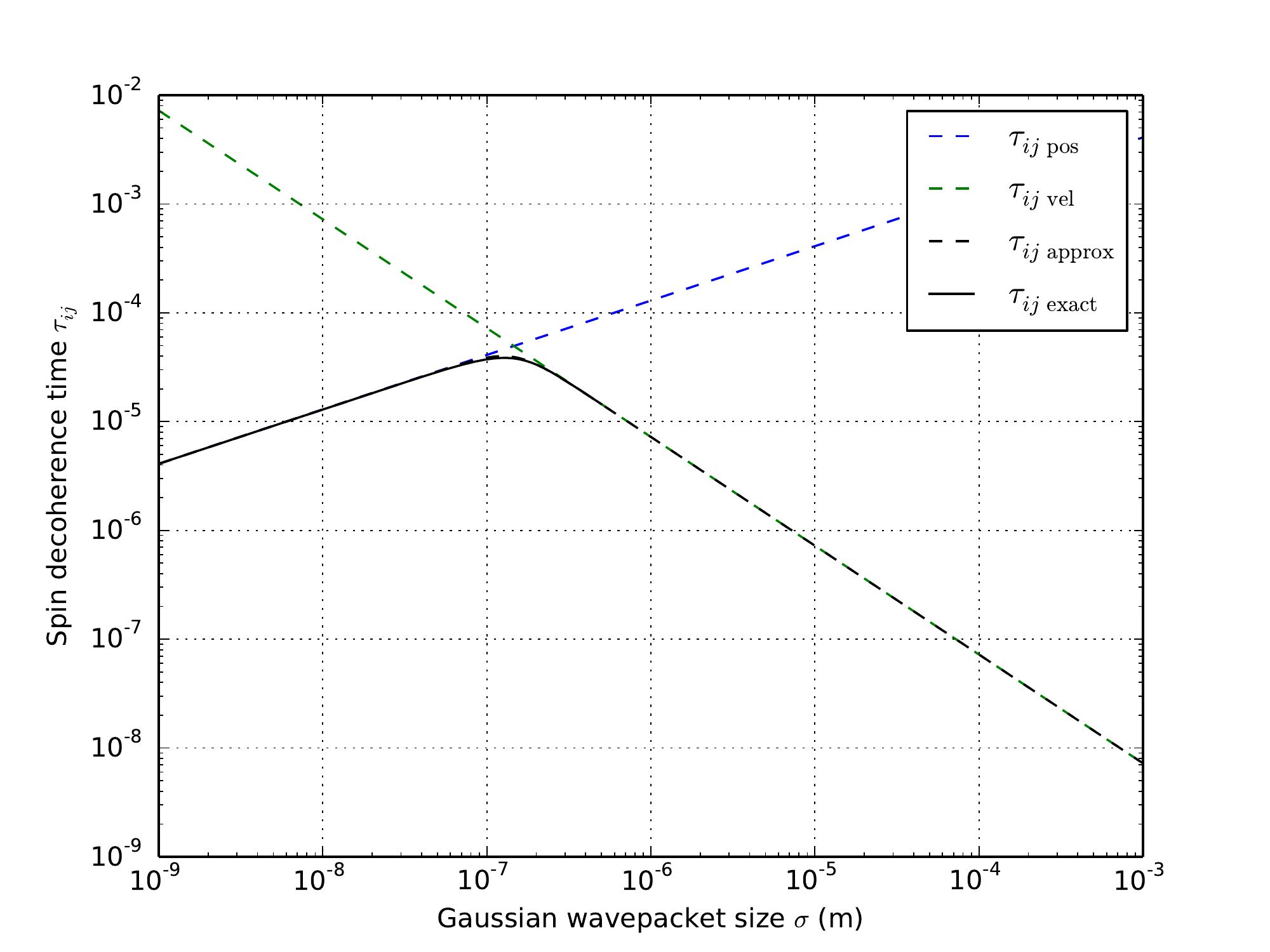}
    \else
        \includegraphics[width=\textwidth]{suppinfofigure1.pdf}
    \fi
    \caption{The computed decoherence time between the $m_F=\pm 1$ states of the $F=1$ hyperfine groundstate of $^{87}\mathrm{Rb}$ in a $125\,\mathrm{Gcm}^{-1}$ magnetic field gradient. At small wavepacket sizes, position separation  dominates the decoherence (equation (\ref{supp:eq:taupos})), whereas at larger wavepacket sizes velocity separation (equation (\ref{supp:eq:tauvel})) dominates. The simple expression (\ref{supp:eq:tauapprox}) gives a good approximation to more complicated exact solution given by (\ref{supp:eq:tauexact})} over the domain, although the latter was not computed over the entire domain of the plot due to numerical overflow.
    \label{supp:fig:approx}
    \end{centering}
\end{figure}

Let's treat this as an unnormalized probability distribution for the wavepackets still being overlapped, and thus define the average separation time $\tau_{ij} \equiv \braket{t}$ as:
\begin{align}
\tau_{ij} & =
\frac{
\int_0^\infty t\left|\braket {\psi_i(t) | \psi_j(t)}\right|\,\mathrm{d}t
}
{
\int_0^\infty \left|\braket {\psi_i(t) | \psi_j(t)}\right|\,\mathrm{d}t
}.\label{supp:eq:taudef}
\end{align}
This results in the following mouthful:
\begin{align}
\tau_{ij\,\mathrm{exact}} & =
\frac{
\sqrt{2\pi}\hbar\exp\left[\eta^2\right]\mathrm{erfc}\left(\sqrt{2}\eta\right)
}
{
m\sigma a_{ij}K_{\frac14}\left(\eta^2\right)
},
\label{supp:eq:tauexact}
\end{align}
where $\mathrm{erfc}$ is the complementary error function, $K_{\frac14}$ is a modified Bessel function of the second kind and \mbox{$\eta = m^2\sigma^3a_{ij}\hbar^{-2}$} is a dimensionless parameter.

The above is not just inconvenient to calculate, in this form it is numerically difficult too, as the exponential overflows easily despite the overall expression being equal to a value that would not overflow. So let's come up with an approximation. There are two regimes in which we can get very simple expressions for $\tau_{ij}$, they are in the limits of small and large wavepacket sizes. For small wavepackets ($\eta \ll 1$) we neglect the second term in (\ref{supp:eq:modoverlap}). This is equivalent to ignoring the velocity difference between the two wavepackets, and so gives us a time constant for the decrease in overlap purely due to separation in position space. Solving the integral (\ref{supp:eq:taudef}) with this approximation gives:
\begin{align}
\tau_{ij\,\mathrm{pos}} & =
\frac{(2\pi^2)^{\frac14}}{2\Gamma(\frac54)}
\sqrt{\frac{\sigma}{a_{ij}}}\label{supp:eq:taupos}
\\
& \approx
1.163
\sqrt{\frac{\sigma}{a_{ij}}}.
\end{align}
Neglecting instead the first term in (\ref{supp:eq:modoverlap}) gives us the time constant for velocity separation, valid when $\eta \gg 1$:
\begin{align}
\tau_{ij\,\mathrm{vel}} & =
\sqrt{\frac2\pi}\frac\hbar{m\sigma a_{ij}}\label{supp:eq:tauvel}
\\
& \approx
0.798
\frac\hbar{m\sigma a_{ij}}.
\end{align}
Unless $\eta\sim1$, one of these time constants will be much smaller than the other, and so the larger can be neglected if this is known to be the case. However, for an approximate expression that is valid for all wavepacket sizes, one can add the cubes of their reciprocals\footnote{If these time constants were for pure exponential decay, we could add their reciprocals directly. If they were for Gaussian decay, we could add the squares of their reciprocals. However one of our time constants is for Gaussian decay and the other is for decay which is quartic in $t$ in the exponent, as per (\ref{supp:eq:timeconstants}), so adding the \emph{cubes} of their reciprocals (3 being close to the geometric mean of 2 and 4), gives a good approximation.}, giving:
\begin{align}
\frac1{\tau_{ij\,\mathrm{approx}}} & = \left[\frac1{\tau^3_{ij\,\mathrm{pos}}} + \frac1{\tau^3_{ij\,\mathrm{vel}}}\right]^{\frac13}.\label{supp:eq:tauapprox}
\end{align}
A plot showing the accuracy of this expression compared to the exact solution is shown in figure~\ref{supp:fig:approx}.

Note that for large wavepacket sizes, the single mode approximation may not be valid: a magnetic field with features smaller than the wavepacket size may induce spin transitions in some parts of the wavepacket but not others. The very short decoherence times shown in figure~\ref{supp:fig:approx} for large wavepackets would therefore be unphysical. If the spin decoherence time is short compared to the timescale of spin transitions, this may unphysically suppress spin transitions via the quantum Zeno effect~\cite{itano1990}. Caution should therefore be exercised not to choose a wavepacket size larger than the range over which the single mode approximation can reasonably be expected to hold.

The simulations in our paper use the thermal wavelength at a particular temperature for the wavepacket size, which is the size of a wavepacket in a thermal gas shortly after a collision---considerably shorter than the time averaged wavepacket size, which is on the order of the mean free path~\cite{busse2009}.

\ifarxiv
\else
    \bibliographystyle{unsrt}
    \bibliography{bibliography}

\begin{thebibliography}{32}%
\makeatletter
\providecommand \@ifxundefined [1]{%
 \@ifx{#1\undefined}
}%
\providecommand \@ifnum [1]{%
 \ifnum #1\expandafter \@firstoftwo
 \else \expandafter \@secondoftwo
 \fi
}%
\providecommand \@ifx [1]{%
 \ifx #1\expandafter \@firstoftwo
 \else \expandafter \@secondoftwo
 \fi
}%
\providecommand \natexlab [1]{#1}%
\providecommand \enquote  [1]{``#1''}%
\providecommand \bibnamefont  [1]{#1}%
\providecommand \bibfnamefont [1]{#1}%
\providecommand \citenamefont [1]{#1}%
\providecommand \href@noop [0]{\@secondoftwo}%
\providecommand \href [0]{\begingroup \@sanitize@url \@href}%
\providecommand \@href[1]{\@@startlink{#1}\@@href}%
\providecommand \@@href[1]{\endgroup#1\@@endlink}%
\providecommand \@sanitize@url [0]{\catcode `\\12\catcode `\$12\catcode
  `\&12\catcode `\#12\catcode `\^12\catcode `\_12\catcode `\%12\relax}%
\providecommand \@@startlink[1]{}%
\providecommand \@@endlink[0]{}%
\providecommand \url  [0]{\begingroup\@sanitize@url \@url }%
\providecommand \@url [1]{\endgroup\@href {#1}{\urlprefix }}%
\providecommand \urlprefix  [0]{URL }%
\providecommand \Eprint [0]{\href }%
\providecommand \doibase [0]{http://dx.doi.org/}%
\providecommand \selectlanguage [0]{\@gobble}%
\providecommand \bibinfo  [0]{\@secondoftwo}%
\providecommand \bibfield  [0]{\@secondoftwo}%
\providecommand \translation [1]{[#1]}%
\providecommand \BibitemOpen [0]{}%
\providecommand \bibitemStop [0]{}%
\providecommand \bibitemNoStop [0]{.\EOS\space}%
\providecommand \EOS [0]{\spacefactor3000\relax}%
\providecommand \BibitemShut  [1]{\csname bibitem#1\endcsname}%
\let\auto@bib@innerbib\@empty
\bibitem [{\citenamefont {Stenholm}(1986)}]{stenholm1986}%
  \BibitemOpen
  \bibfield  {author} {\bibinfo {author} {\bibfnamefont {S.}~\bibnamefont
  {Stenholm}},\ }\href {\doibase 10.1103/RevModPhys.58.699} {\bibfield
  {journal} {\bibinfo  {journal} {Rev. Mod. Phys.}\ }\textbf {\bibinfo {volume}
  {58}},\ \bibinfo {pages} {699} (\bibinfo {year} {1986})}\BibitemShut
  {NoStop}%
\bibitem [{\citenamefont {Javanainen}(1992)}]{javanainen1992}%
  \BibitemOpen
  \bibfield  {author} {\bibinfo {author} {\bibfnamefont {J.}~\bibnamefont
  {Javanainen}},\ }\href {\doibase 10.1103/PhysRevA.46.5819} {\bibfield
  {journal} {\bibinfo  {journal} {Phys. Rev. A}\ }\textbf {\bibinfo {volume}
  {46}},\ \bibinfo {pages} {5819} (\bibinfo {year} {1992})}\BibitemShut
  {NoStop}%
\bibitem [{\citenamefont {Dalibard}\ and\ \citenamefont
  {Cohen-Tannoudji}(1989)}]{dalibard89}%
  \BibitemOpen
  \bibfield  {author} {\bibinfo {author} {\bibfnamefont {J.}~\bibnamefont
  {Dalibard}}\ and\ \bibinfo {author} {\bibfnamefont {C.}~\bibnamefont
  {Cohen-Tannoudji}},\ }\href {\doibase 10.1364/JOSAB.6.002023} {\bibfield
  {journal} {\bibinfo  {journal} {J. Opt. Soc. Am. B}\ }\textbf {\bibinfo
  {volume} {6}},\ \bibinfo {pages} {2023} (\bibinfo {year} {1989})}\BibitemShut
  {NoStop}%
\bibitem [{\citenamefont {Domokos}\ \emph {et~al.}(2001)\citenamefont
  {Domokos}, \citenamefont {Horak},\ and\ \citenamefont
  {Ritsch}}]{domokos2001}%
  \BibitemOpen
  \bibfield  {author} {\bibinfo {author} {\bibfnamefont {P.}~\bibnamefont
  {Domokos}}, \bibinfo {author} {\bibfnamefont {P.}~\bibnamefont {Horak}}, \
  and\ \bibinfo {author} {\bibfnamefont {H.}~\bibnamefont {Ritsch}},\ }\href
  {http://stacks.iop.org/0953-4075/34/i=2/a=306} {\bibfield  {journal}
  {\bibinfo  {journal} {Journal of Physics B: Atomic, Molecular and Optical
  Physics}\ }\textbf {\bibinfo {volume} {34}},\ \bibinfo {pages} {187}
  (\bibinfo {year} {2001})}\BibitemShut {NoStop}%
\bibitem [{\citenamefont {Wade}\ \emph {et~al.}(2011)\citenamefont {Wade},
  \citenamefont {Baillie},\ and\ \citenamefont {Blakie}}]{Wade2011}%
  \BibitemOpen
  \bibfield  {author} {\bibinfo {author} {\bibfnamefont {A.~C.~J.}\
  \bibnamefont {Wade}}, \bibinfo {author} {\bibfnamefont {D.}~\bibnamefont
  {Baillie}}, \ and\ \bibinfo {author} {\bibfnamefont {P.~B.}\ \bibnamefont
  {Blakie}},\ }\href {\doibase 10.1103/PhysRevA.84.023612} {\bibfield
  {journal} {\bibinfo  {journal} {Phys. Rev. A}\ }\textbf {\bibinfo {volume}
  {84}},\ \bibinfo {pages} {023612} (\bibinfo {year} {2011})}\BibitemShut
  {NoStop}%
\bibitem [{\citenamefont {Busse}\ and\ \citenamefont
  {Hornberger}(2009)}]{busse2009}%
  \BibitemOpen
  \bibfield  {author} {\bibinfo {author} {\bibfnamefont {M.}~\bibnamefont
  {Busse}}\ and\ \bibinfo {author} {\bibfnamefont {K.}~\bibnamefont
  {Hornberger}},\ }\href {http://stacks.iop.org/1751-8121/42/i=36/a=362001}
  {\bibfield  {journal} {\bibinfo  {journal} {Journal of Physics A:
  Mathematical and Theoretical}\ }\textbf {\bibinfo {volume} {42}},\ \bibinfo
  {pages} {362001} (\bibinfo {year} {2009})}\BibitemShut {NoStop}%
\bibitem [{\citenamefont {Ehrenfest}(1927)}]{Ehrenfest1927}%
  \BibitemOpen
  \bibfield  {author} {\bibinfo {author} {\bibfnamefont {P.}~\bibnamefont
  {Ehrenfest}},\ }\href {\doibase 10.1007/BF01329203} {\bibfield  {journal}
  {\bibinfo  {journal} {Zeitschrift f{\"u}r Physik}\ }\textbf {\bibinfo
  {volume} {45}},\ \bibinfo {pages} {455} (\bibinfo {year} {1927})}\BibitemShut
  {NoStop}%
\bibitem [{\citenamefont {Salomon}\ \emph {et~al.}(1990)\citenamefont
  {Salomon}, \citenamefont {Dalibard}, \citenamefont {Phillips}, \citenamefont
  {Clairon},\ and\ \citenamefont {Guellati}}]{salomon1990}%
  \BibitemOpen
  \bibfield  {author} {\bibinfo {author} {\bibfnamefont {C.}~\bibnamefont
  {Salomon}}, \bibinfo {author} {\bibfnamefont {J.}~\bibnamefont {Dalibard}},
  \bibinfo {author} {\bibfnamefont {W.~D.}\ \bibnamefont {Phillips}}, \bibinfo
  {author} {\bibfnamefont {A.}~\bibnamefont {Clairon}}, \ and\ \bibinfo
  {author} {\bibfnamefont {S.}~\bibnamefont {Guellati}},\ }\href
  {http://stacks.iop.org/0295-5075/12/i=8/a=003} {\bibfield  {journal}
  {\bibinfo  {journal} {EPL (Europhysics Letters)}\ }\textbf {\bibinfo {volume}
  {12}},\ \bibinfo {pages} {683} (\bibinfo {year} {1990})}\BibitemShut
  {NoStop}%
\bibitem [{\citenamefont {Pritchard}(1983)}]{pritchard1983}%
  \BibitemOpen
  \bibfield  {author} {\bibinfo {author} {\bibfnamefont {D.~E.}\ \bibnamefont
  {Pritchard}},\ }\href {\doibase 10.1103/PhysRevLett.51.1336} {\bibfield
  {journal} {\bibinfo  {journal} {Phys. Rev. Lett.}\ }\textbf {\bibinfo
  {volume} {51}},\ \bibinfo {pages} {1336} (\bibinfo {year}
  {1983})}\BibitemShut {NoStop}%
\bibitem [{\citenamefont {Davis}\ \emph {et~al.}(1995)\citenamefont {Davis},
  \citenamefont {Mewes}, \citenamefont {Andrews}, \citenamefont {van Druten},
  \citenamefont {Durfee}, \citenamefont {Kurn},\ and\ \citenamefont
  {Ketterle}}]{davis1995}%
  \BibitemOpen
  \bibfield  {author} {\bibinfo {author} {\bibfnamefont {K.~B.}\ \bibnamefont
  {Davis}}, \bibinfo {author} {\bibfnamefont {M.~O.}\ \bibnamefont {Mewes}},
  \bibinfo {author} {\bibfnamefont {M.~R.}\ \bibnamefont {Andrews}}, \bibinfo
  {author} {\bibfnamefont {N.~J.}\ \bibnamefont {van Druten}}, \bibinfo
  {author} {\bibfnamefont {D.~S.}\ \bibnamefont {Durfee}}, \bibinfo {author}
  {\bibfnamefont {D.~M.}\ \bibnamefont {Kurn}}, \ and\ \bibinfo {author}
  {\bibfnamefont {W.}~\bibnamefont {Ketterle}},\ }\href {\doibase
  10.1103/PhysRevLett.75.3969} {\bibfield  {journal} {\bibinfo  {journal}
  {Phys. Rev. Lett.}\ }\textbf {\bibinfo {volume} {75}},\ \bibinfo {pages}
  {3969} (\bibinfo {year} {1995})}\BibitemShut {NoStop}%
\bibitem [{\citenamefont {Anderson}\ \emph {et~al.}(1995)\citenamefont
  {Anderson}, \citenamefont {Ensher}, \citenamefont {Matthews}, \citenamefont
  {Wieman},\ and\ \citenamefont {Cornell}}]{anderson1995}%
  \BibitemOpen
  \bibfield  {author} {\bibinfo {author} {\bibfnamefont {M.~H.}\ \bibnamefont
  {Anderson}}, \bibinfo {author} {\bibfnamefont {J.~R.}\ \bibnamefont
  {Ensher}}, \bibinfo {author} {\bibfnamefont {M.~R.}\ \bibnamefont
  {Matthews}}, \bibinfo {author} {\bibfnamefont {C.~E.}\ \bibnamefont
  {Wieman}}, \ and\ \bibinfo {author} {\bibfnamefont {E.~A.}\ \bibnamefont
  {Cornell}},\ }\href {\doibase 10.1126/science.269.5221.198} {\bibfield
  {journal} {\bibinfo  {journal} {Science}\ }\textbf {\bibinfo {volume}
  {269}},\ \bibinfo {pages} {198} (\bibinfo {year} {1995})}\BibitemShut
  {NoStop}%
\bibitem [{\citenamefont {Majorana}(1932)}]{majorana1932}%
  \BibitemOpen
  \bibfield  {author} {\bibinfo {author} {\bibfnamefont {E.}~\bibnamefont
  {Majorana}},\ }\href@noop {} {\bibfield  {journal} {\bibinfo  {journal} {Il
  Nuovo Cimento}\ }\textbf {\bibinfo {volume} {9}},\ \bibinfo {pages} {43}
  (\bibinfo {year} {1932})}\BibitemShut {NoStop}%
\bibitem [{\citenamefont {Gerlach}\ and\ \citenamefont
  {Stern}(1922)}]{Gerlach1922}%
  \BibitemOpen
  \bibfield  {author} {\bibinfo {author} {\bibfnamefont {W.}~\bibnamefont
  {Gerlach}}\ and\ \bibinfo {author} {\bibfnamefont {O.}~\bibnamefont
  {Stern}},\ }\href {\doibase 10.1007/BF01326984} {\bibfield  {journal}
  {\bibinfo  {journal} {Zeitschrift f{\"u}r Physik}\ }\textbf {\bibinfo
  {volume} {9}},\ \bibinfo {pages} {353} (\bibinfo {year} {1922})}\BibitemShut
  {NoStop}%
\bibitem [{\citenamefont {Carmichael}(1993)}]{Carmichael1993}%
  \BibitemOpen
  \bibfield  {author} {\bibinfo {author} {\bibfnamefont {H.~J.}\ \bibnamefont
  {Carmichael}},\ }\href {\doibase 10.1103/PhysRevLett.70.2273} {\bibfield
  {journal} {\bibinfo  {journal} {Phys. Rev. Lett.}\ }\textbf {\bibinfo
  {volume} {70}},\ \bibinfo {pages} {2273} (\bibinfo {year}
  {1993})}\BibitemShut {NoStop}%
\bibitem [{\citenamefont {M{\o}lmer}\ \emph {et~al.}(1993)\citenamefont
  {M{\o}lmer}, \citenamefont {Castin},\ and\ \citenamefont
  {Dalibard}}]{molmer1993}%
  \BibitemOpen
  \bibfield  {author} {\bibinfo {author} {\bibfnamefont {K.}~\bibnamefont
  {M{\o}lmer}}, \bibinfo {author} {\bibfnamefont {Y.}~\bibnamefont {Castin}}, \
  and\ \bibinfo {author} {\bibfnamefont {J.}~\bibnamefont {Dalibard}},\ }\href
  {\doibase 10.1364/JOSAB.10.000524} {\bibfield  {journal} {\bibinfo  {journal}
  {J. Opt. Soc. Am. B}\ }\textbf {\bibinfo {volume} {10}},\ \bibinfo {pages}
  {524} (\bibinfo {year} {1993})}\BibitemShut {NoStop}%
\bibitem [{\citenamefont {Teich}\ and\ \citenamefont
  {Mahler}(1992)}]{teich1992}%
  \BibitemOpen
  \bibfield  {author} {\bibinfo {author} {\bibfnamefont {W.~G.}\ \bibnamefont
  {Teich}}\ and\ \bibinfo {author} {\bibfnamefont {G.}~\bibnamefont {Mahler}},\
  }\href {\doibase 10.1103/PhysRevA.45.3300} {\bibfield  {journal} {\bibinfo
  {journal} {Phys. Rev. A}\ }\textbf {\bibinfo {volume} {45}},\ \bibinfo
  {pages} {3300} (\bibinfo {year} {1992})}\BibitemShut {NoStop}%
\bibitem [{\citenamefont {Gisin}\ and\ \citenamefont
  {Percival}(1992)}]{gisin1992}%
  \BibitemOpen
  \bibfield  {author} {\bibinfo {author} {\bibfnamefont {N.}~\bibnamefont
  {Gisin}}\ and\ \bibinfo {author} {\bibfnamefont {I.~C.}\ \bibnamefont
  {Percival}},\ }\href {\doibase
  http://dx.doi.org/10.1016/0375-9601(92)90264-M} {\bibfield  {journal}
  {\bibinfo  {journal} {Physics Letters A}\ }\textbf {\bibinfo {volume}
  {167}},\ \bibinfo {pages} {315 } (\bibinfo {year} {1992})}\BibitemShut
  {NoStop}%
\bibitem [{\citenamefont {Wiseman}(1996)}]{wiseman1996}%
  \BibitemOpen
  \bibfield  {author} {\bibinfo {author} {\bibfnamefont {H.~M.}\ \bibnamefont
  {Wiseman}},\ }\href {http://stacks.iop.org/1355-5111/8/i=1/a=015} {\bibfield
  {journal} {\bibinfo  {journal} {Quantum and Semiclassical Optics: Journal of
  the European Optical Society Part B}\ }\textbf {\bibinfo {volume} {8}},\
  \bibinfo {pages} {205} (\bibinfo {year} {1996})}\BibitemShut {NoStop}%
\bibitem [{\citenamefont {Plenio}\ and\ \citenamefont
  {Knight}(1998)}]{plenio1998}%
  \BibitemOpen
  \bibfield  {author} {\bibinfo {author} {\bibfnamefont {M.~B.}\ \bibnamefont
  {Plenio}}\ and\ \bibinfo {author} {\bibfnamefont {P.~L.}\ \bibnamefont
  {Knight}},\ }\href {\doibase 10.1103/RevModPhys.70.101} {\bibfield  {journal}
  {\bibinfo  {journal} {Rev. Mod. Phys.}\ }\textbf {\bibinfo {volume} {70}},\
  \bibinfo {pages} {101} (\bibinfo {year} {1998})}\BibitemShut {NoStop}%
\bibitem [{\citenamefont {Bouwmeester}\ and\ \citenamefont
  {Nienhuis}(1996)}]{Bouwmeester1996}%
  \BibitemOpen
  \bibfield  {author} {\bibinfo {author} {\bibfnamefont {D.}~\bibnamefont
  {Bouwmeester}}\ and\ \bibinfo {author} {\bibfnamefont {G.}~\bibnamefont
  {Nienhuis}},\ }\href {http://stacks.iop.org/1355-5111/8/i=1/a=020} {\bibfield
   {journal} {\bibinfo  {journal} {Quantum and Semiclassical Optics: Journal of
  the European Optical Society Part B}\ }\textbf {\bibinfo {volume} {8}},\
  \bibinfo {pages} {277} (\bibinfo {year} {1996})}\BibitemShut {NoStop}%
\bibitem [{\citenamefont {Zurek}(1991)}]{zurek1991}%
  \BibitemOpen
  \bibfield  {author} {\bibinfo {author} {\bibfnamefont {W.~H.}\ \bibnamefont
  {Zurek}},\ }\href {\doibase 10.1063/1.881293} {\bibfield  {journal} {\bibinfo
   {journal} {Physics Today}\ }\textbf {\bibinfo {volume} {44}},\ \bibinfo
  {pages} {36} (\bibinfo {year} {1991})}\BibitemShut {NoStop}%
\bibitem [{\citenamefont {Zurek}(2003)}]{zurek2003}%
  \BibitemOpen
  \bibfield  {author} {\bibinfo {author} {\bibfnamefont {W.~H.}\ \bibnamefont
  {Zurek}},\ }\href {\doibase 10.1103/RevModPhys.75.715} {\bibfield  {journal}
  {\bibinfo  {journal} {Rev. Mod. Phys.}\ }\textbf {\bibinfo {volume} {75}},\
  \bibinfo {pages} {715} (\bibinfo {year} {2003})}\BibitemShut {NoStop}%
\bibitem [{\citenamefont {Schlosshauer}(2007)}]{schlosshauer2007}%
  \BibitemOpen
  \bibfield  {author} {\bibinfo {author} {\bibfnamefont {M.}~\bibnamefont
  {Schlosshauer}},\ }\href {http://books.google.com.au/books?id=1qrJUS5zNbEC}
  {\emph {\bibinfo {title} {Decoherence: And the Quantum-To-Classical
  Transition}}},\ The Frontiers Collection\ (\bibinfo  {publisher} {Springer},\
  \bibinfo {year} {2007})\BibitemShut {NoStop}%
\bibitem [{\citenamefont {Raithel}\ and\ \citenamefont
  {Morrow}(2006)}]{Raithel2006}%
  \BibitemOpen
  \bibfield  {author} {\bibinfo {author} {\bibfnamefont {G.}~\bibnamefont
  {Raithel}}\ and\ \bibinfo {author} {\bibfnamefont {N.}~\bibnamefont
  {Morrow}},\ }\href@noop {} {\bibfield  {journal} {\bibinfo  {journal}
  {Advances in Atomic, Molecular and Optical Physics}\ }\textbf {\bibinfo
  {volume} {53}},\ \bibinfo {pages} {187} (\bibinfo {year} {2006})}\BibitemShut
  {NoStop}%
\bibitem [{\citenamefont {Walls}\ and\ \citenamefont
  {Milburn}(1985)}]{walls1985}%
  \BibitemOpen
  \bibfield  {author} {\bibinfo {author} {\bibfnamefont {D.~F.}\ \bibnamefont
  {Walls}}\ and\ \bibinfo {author} {\bibfnamefont {G.~J.}\ \bibnamefont
  {Milburn}},\ }\href {\doibase 10.1103/PhysRevA.31.2403} {\bibfield  {journal}
  {\bibinfo  {journal} {Phys. Rev. A}\ }\textbf {\bibinfo {volume} {31}},\
  \bibinfo {pages} {2403} (\bibinfo {year} {1985})}\BibitemShut {NoStop}%
\bibitem [{\citenamefont {Moy}\ \emph {et~al.}(1999)\citenamefont {Moy},
  \citenamefont {Hope},\ and\ \citenamefont {Savage}}]{moy1999}%
  \BibitemOpen
  \bibfield  {author} {\bibinfo {author} {\bibfnamefont {G.~M.}\ \bibnamefont
  {Moy}}, \bibinfo {author} {\bibfnamefont {J.~J.}\ \bibnamefont {Hope}}, \
  and\ \bibinfo {author} {\bibfnamefont {C.~M.}\ \bibnamefont {Savage}},\
  }\href {\doibase 10.1103/PhysRevA.59.667} {\bibfield  {journal} {\bibinfo
  {journal} {Phys. Rev. A}\ }\textbf {\bibinfo {volume} {59}},\ \bibinfo
  {pages} {667} (\bibinfo {year} {1999})}\BibitemShut {NoStop}%
\bibitem [{\citenamefont {Walborn}\ \emph {et~al.}(2002)\citenamefont
  {Walborn}, \citenamefont {Terra~Cunha}, \citenamefont {P\'adua},\ and\
  \citenamefont {Monken}}]{Walborn2002}%
  \BibitemOpen
  \bibfield  {author} {\bibinfo {author} {\bibfnamefont {S.~P.}\ \bibnamefont
  {Walborn}}, \bibinfo {author} {\bibfnamefont {M.~O.}\ \bibnamefont
  {Terra~Cunha}}, \bibinfo {author} {\bibfnamefont {S.}~\bibnamefont
  {P\'adua}}, \ and\ \bibinfo {author} {\bibfnamefont {C.~H.}\ \bibnamefont
  {Monken}},\ }\href {\doibase 10.1103/PhysRevA.65.033818} {\bibfield
  {journal} {\bibinfo  {journal} {Phys. Rev. A}\ }\textbf {\bibinfo {volume}
  {65}},\ \bibinfo {pages} {033818} (\bibinfo {year} {2002})}\BibitemShut
  {NoStop}%
\bibitem [{\citenamefont {Bergquist}\ \emph {et~al.}(1986)\citenamefont
  {Bergquist}, \citenamefont {Hulet}, \citenamefont {Itano},\ and\
  \citenamefont {Wineland}}]{Bergquist1986}%
  \BibitemOpen
  \bibfield  {author} {\bibinfo {author} {\bibfnamefont {J.~C.}\ \bibnamefont
  {Bergquist}}, \bibinfo {author} {\bibfnamefont {R.~G.}\ \bibnamefont
  {Hulet}}, \bibinfo {author} {\bibfnamefont {W.~M.}\ \bibnamefont {Itano}}, \
  and\ \bibinfo {author} {\bibfnamefont {D.~J.}\ \bibnamefont {Wineland}},\
  }\href {\doibase 10.1103/PhysRevLett.57.1699} {\bibfield  {journal} {\bibinfo
   {journal} {Phys. Rev. Lett.}\ }\textbf {\bibinfo {volume} {57}},\ \bibinfo
  {pages} {1699} (\bibinfo {year} {1986})}\BibitemShut {NoStop}%
\bibitem [{\citenamefont {Sauter}\ \emph {et~al.}(1986)\citenamefont {Sauter},
  \citenamefont {Neuhauser}, \citenamefont {Blatt},\ and\ \citenamefont
  {Toschek}}]{Sauter1986}%
  \BibitemOpen
  \bibfield  {author} {\bibinfo {author} {\bibfnamefont {T.}~\bibnamefont
  {Sauter}}, \bibinfo {author} {\bibfnamefont {W.}~\bibnamefont {Neuhauser}},
  \bibinfo {author} {\bibfnamefont {R.}~\bibnamefont {Blatt}}, \ and\ \bibinfo
  {author} {\bibfnamefont {P.~E.}\ \bibnamefont {Toschek}},\ }\href {\doibase
  10.1103/PhysRevLett.57.1696} {\bibfield  {journal} {\bibinfo  {journal}
  {Phys. Rev. Lett.}\ }\textbf {\bibinfo {volume} {57}},\ \bibinfo {pages}
  {1696} (\bibinfo {year} {1986})}\BibitemShut {NoStop}%
\bibitem [{\citenamefont {Dennis}\ \emph {et~al.}(2013)\citenamefont {Dennis},
  \citenamefont {Hope},\ and\ \citenamefont {Johnsson}}]{Dennis2013}%
  \BibitemOpen
  \bibfield  {author} {\bibinfo {author} {\bibfnamefont {G.~R.}\ \bibnamefont
  {Dennis}}, \bibinfo {author} {\bibfnamefont {J.~J.}\ \bibnamefont {Hope}}, \
  and\ \bibinfo {author} {\bibfnamefont {M.~T.}\ \bibnamefont {Johnsson}},\
  }\href {\doibase http://dx.doi.org/10.1016/j.cpc.2012.08.016} {\bibfield
  {journal} {\bibinfo  {journal} {Computer Physics Communications}\ }\textbf
  {\bibinfo {volume} {184}},\ \bibinfo {pages} {201 } (\bibinfo {year}
  {2013})}\BibitemShut {NoStop}%
\bibitem [{\citenamefont {Aaronson}(2005)}]{Aaronson2005}%
  \BibitemOpen
  \bibfield  {author} {\bibinfo {author} {\bibfnamefont {S.}~\bibnamefont
  {Aaronson}},\ }\href {\doibase 10.1103/PhysRevA.71.032325} {\bibfield
  {journal} {\bibinfo  {journal} {Phys. Rev. A}\ }\textbf {\bibinfo {volume}
  {71}},\ \bibinfo {pages} {032325} (\bibinfo {year} {2005})}\BibitemShut
  {NoStop}%
\bibitem [{\citenamefont {Itano}\ \emph {et~al.}(1990)\citenamefont {Itano},
  \citenamefont {Heinzen}, \citenamefont {Bollinger},\ and\ \citenamefont
  {Wineland}}]{itano1990}%
  \BibitemOpen
  \bibfield  {author} {\bibinfo {author} {\bibfnamefont {W.~M.}\ \bibnamefont
  {Itano}}, \bibinfo {author} {\bibfnamefont {D.~J.}\ \bibnamefont {Heinzen}},
  \bibinfo {author} {\bibfnamefont {J.~J.}\ \bibnamefont {Bollinger}}, \ and\
  \bibinfo {author} {\bibfnamefont {D.~J.}\ \bibnamefont {Wineland}},\ }\href
  {\doibase 10.1103/PhysRevA.41.2295} {\bibfield  {journal} {\bibinfo
  {journal} {Phys. Rev. A}\ }\textbf {\bibinfo {volume} {41}},\ \bibinfo
  {pages} {2295} (\bibinfo {year} {1990})}\BibitemShut {NoStop}%
\end{thebibliography}%
\fi

\end{document}

\fi
\end{document}